\documentclass[epj]{svjour}
\usepackage{latexsym}
\usepackage{graphics}
\usepackage{epsfig}
\usepackage{makeidx}
\usepackage{graphicx}
\usepackage{epstopdf}
\begin{document}
\authorrunning{M.L. Bertotti and G. Modanese}
\titlerunning{A Family of Taxation and Redistribution Models}
\title{Exploiting the flexibility of a family of models for taxation and redistribution}
%Title
%%%%%\subtitle{Do you have a subtitle?\\ If so, write it here}
\author{Maria Letizia Bertotti%\inst{1} 
\and 
Giovanni Modanese
%\inst{2}% etc
% \thanks is optional - remove next line if not needed
%%%%%\thanks{\emph{Present address:} Insert the address here if needed}%
}                     % Do not remove
\offprints{}          % Insert a name or remove this line
\institute{Faculty of Science and Technology,
   Free University of Bozen-Bolzano,
   Piazza Universit\`a 5, 39100 Bolzano, ITALY
   %\email{MariaLetizia.Bertotti@unibz.it}
   %\and 
   %\email{Giovanni.Modanese@unibz.it}
   }
%%%%
%
\date{Received: date / Revised version: date}
% The correct dates will be entered by Springer
%
\abstract{
We discuss a family of models expressed by nonlinear differential equation systems
describing closed market societies in the presence of taxation and redistribution.
We focus in particular on three example models obtained in correspondence to different parameter 
choices. 
We analyse the influence of the various choices on the long time shape of the income distribution.
Several simulations suggest that behavioral heterogeneity among the individuals plays a definite role 
in the formation of fat tails of the asymptotic stationary distributions.
This is in agreement with results found with different approaches and techniques.
We also show that an excellent fit for the computational outputs of our models is provided by
the $\kappa$-generalized distribution introduced by Kaniadakis in \cite{Kan}.
} %end of abstract
\maketitle
%

%I put myself the following few commands
\def\pb{\, .}
\def\vb{\, ,}

%%%%%%%%%%%%%%%%%%%%%%%%%%%%%%%%%%%%%%%%%%%%%
\section{Introduction}
\label{intro}
%%%%%%%%%%%%%%%%%%%%%%%%%%%%%%%%%%%%%%%%%%%%%

Several papers have been published,
especially during the last decade, 
which aim at explaining
some universal features of income and wealth distributions
with the tools of statistical mechanics 
and kinetic theory.
We recall for example
\cite{DraYak}, \cite{ChaCha},
\cite{ChaFigMouRib},
\cite{ChattCha},
\cite{DurMatTos},
and refer to \cite{Yak} or \cite{YakRos} for a 
detailed review and bibliography.
The models 
described
in most of these papers 
typically concern
closed market societies in which
wealth exchanges 
obeying some given rule and
involving pairs of agents
take place. The challenge is to derive from 
the knowledge of
the microscopic interactions
the emergence of collective 
patterns
similar to those observed in the real world. 
The techniques more frequently employed
take advantage of
today's
computer power and
include
Monte Carlo methods and agent based 
or numerical simulations.

In a recent work \cite{BerMod} 
belonging to this research line
the authors of this note
investigated
a family of models for the taxation and redistribution process,
formulated within a general framework
proposed in \cite{Ber}.
The models at issue are
expressed by systems of nonlinear ordinary differential equations,
which enjoy a great descriptive flexibility
thanks to the presence of various parameters.
The equation systems are well beyond analytic solution.
Nevertheless, many computational simulations
give evidence of the existence of asymptotic stationary  distributions 
whose higher income sections
exhibit,
at least for some choices of the model parameters,
a power law like decreasing behavior.
Together with other aspects discussed in \cite{BerMod} and \cite{Ber},
this fact encourages to pursue the study of these models.
Indeed, as well known 
after Pareto observation 
of more than a century ago \cite{Par},
often in real economies
the wealth distribution of the richest people 
follows a power law pattern.\footnote{\ The words income and wealth are used in this paper to designate the same concept.}

In particular, the question arises of what is within the framework and the models 
introduced in \cite{BerMod} 
the cause 
of the detected
decreasing
behavior.
This is the 
first question addressed in this paper.
A mechanism responsible for the
appearence of a power law tail was recognized in 
\cite{ChattChakrManna} and \cite{ChattCha} 
for a class of kinetic wealth exchange models
to be
the so called saving propensity,
a parameter expressing the attitude of agents to retain,
when trading, 
a certain fraction of their wealth.
An analysis of the influence on the shape of the distribution
of homogeneous and heterogeneous (i.e., equal or diversified) saving propensity
among agents of a population can be found 
also in \cite{PatHeinCha}, where various 
kinetic wealth exchange
models of both types are surveyed.
Furthermore, in \cite{MatTos} a rigorous analytic investigation is carried out
for a class of conservative economy models described by a (continuous) Boltzmann-like equation,
which proves in particular that 
randomness in the microscopic trade and saving propensity
play a significant role in
the large wealth behavior of the steady distributions.

Taking advantage of the high flexibility 
which characterizes the models in \cite{BerMod},
we give account below of several numerical simulations 
run in correspondence to 
various choices of  
parameters.
Modulo the different model formulation, these simulations
suggest,
in substantial agreement with the above mentioned references, 
that an heterogeneous saving propensity 
is a crucial ingredient for the fatness of the large wealth part  of the stationary distribution.

A second point
in this paper concerns the attempt to find the analytical expression
of a multi-parameter function with which to fit 
the equilibrium distributions obtained 
via numerical simulations.
Various possibilities suggested in the literature
are briefly recalled in Section \ref{reasonable fit}. A comparison among them
singles out
the $\kappa$-generalized distribution introduced and discussed in
\cite{Kan},
\cite{CleGalKan} and
\cite{CleDiMGalKan} 
as 
especially suitable to fit at a time the
computational outputs relative to 
the poorest, average and richest income values
of the asymptotic distributions of our models.

\smallskip

The plan of the paper is as follows.
In Section \ref{fram and mod} we recall the general framework introduced in \cite{Ber} and
discuss a family of models which enjoy minor changes with respect to the one in 
\cite{BerMod}.
In Section \ref{diff cho} we particularize the models, by choosing three specific sets of values for 
their parameters and we focus on the effect
produced by
the different choices on the shape of the 
asymptotic income distributions. 
In the same section
we also review and shortly compare some 
other wealth exchange models
discussed in the econophysics literature.
In
Section 
\ref{reasonable fit} 
we extend the attention to
the lower income parts
of the asymptotic income distributions found
and we show, for each of the models
of the previous section,
how the \textquotedblleft numerical\textquotedblright distributions fit  with 
suitably tuned
$\kappa$-generalized distributions.
Some remarks on reversibility
can be found in Section \ref{remarks reversibility}. 
In Section  \ref{conclusions}
the results are summarized and discussed within a critical 
perspective.

%%%%%%%%% %%%%%%%%% %%%%%%%%% %%%%%%%%% %%%%%%%%%
\section{The general framework and a model family}
\label{fram and mod}
%%%%%%%%% %%%%%%%%% %%%%%%%%% %%%%%%%%% %%%%%%%%%

Consider a population of individuals divided into a finite number $n$ of classes,
each one characterized by its 
average income.
Let $r_1,r_2 \ldots r_n$
denote the average incomes
of the $n$ classes, ordered so that $r_1\le r_2 \le \ldots \le\ r_n$,
and let $x_i(t)$, where
$x_i : {\bf R} \to [0,+\infty)$ for $i \in \Gamma_n = \{ 1,2, ..., n \}$, denote the fraction at time $t$
of individuals belonging to the $i$-th class.
In the following, the indices $i, j, h, k$, etc. always belong to $\Gamma_n$, if no differently stated.

Assume that pairwise interactions of economic nature, subjected to taxation, take place.
And call $S$ the fixed amount of money that people may exchange during their interactions.

Any time an individual of the $h$-th class has to pay a quantity $S$ to an individual of the $k$-th class, 
this one in turn has to pay some tax corresponding to a percentage of what he is receiving.
This tax is quantified as $S \, \tau$, with the tax rate $\tau = \tau_k \le 1$ depending
in general on the class of the earning individual. Since 
the quantity $S \, \tau$ goes to the government, which
is supposed to use the money collected through taxation to
provide welfare services for the population, we 
interpret the welfare provision as an income redistribution.
Ignoring the passages 
to and from the government,
we 
adopt the following equivalent mechanism as the mover of the dynamics:
in correspondence to any interaction between an $h$-individual and a $k$-individual, where the one who has to pay $S$ to the 
other one is the $h$-individual,
this pays to the $k$-individual a quantity 
$S \, (1 - \tau)$ and he pays as well a quantity $S \, \tau$, which is divided among
all $j$-individuals for $j \ne n$.\footnote{\ The reason why individuals of the $n$-th class constitute an exception is a technical one:
if an individual of the $n$-th class would receive some money, 
the possibility would arise 
for him to advance to a higher class, which is impossible.}
Accordingly, the effect of taxation and redistribution 
is equivalent to the effect of a quantity of interactions 
between the $h$-individual and each one of the $j$-individuals for $j \ne n$,
which are
\textquotedblleft induced\textquotedblright by the effective $h$-$k$ interaction.
To fix notations,
we
may distinguish between 
{\it {direct}} interactions
($h$-$k$) and 
{\it {indirect}} interactions
($h$-$j$ for $j \ne n$).

Any direct or indirect economical interaction
yields as a consequence a possible  slight increase
or slight decrease of
the income of individuals.

To translate into mathematical terms all that, we introduce

- the (table of the) {\slshape interaction rates}
$$
\eta_{hk} \in [0,+\infty)  \vb
$$
expressing the number of effective encounters per unit time between
individuals of the $h$-th class and individuals of the $k$-th class;

- the (tables of the) {\slshape direct transition probability densities} 
$$
C_{hk}^i \in [0,+\infty)  \vb
$$
satisfying for any fixed $h$ and $k$
$$
\sum_{i=1}^n C_{hk}^i = 1  \vb
$$
which 
express the probability density that an individual of the $h$-th class 
will belong to the 
$i$-th class after a direct interaction with an
individual of the $k$-th class;

- the (tables of the) {\slshape indirect transition variation densities} 
$$
T_{[hk]}^i  \, :   {\bf R}^n \to {\bf R}  \vb
$$
where the $T_{[hk]}^i(x)$ with $x=(x_1, ..., x_n) \in {\bf R}^n$ are continuous functions, satisfying, for any fixed $h$, $k$ and
$x  \in {\bf R}^n$ 
$$
\sum_{i=1}^n T_{[hk]}^i(x) = 0 \pb
$$
These functions account for the indirect interactions and express the 
variation density in the $i$-th class
due to an interaction between an individual of the $h$-th class
with an individual of the $k$-th class.

Chose for simplicity all the interactions rates $\eta_{hk}$ to be equal to $1$,
which corresponds to assuming that all the encounters between two individuals occur
with the same frequency, independently of the classes to which the two belong.

Then, the evolution of the class populations $x_i(t)$ 
is governed by the following differential equations, 
in which the contribution of both direct and indirect interactions is present:
\begin{equation}
{{d x_i} \over {d t}} =  
\sum_{h=1}^n \sum_{k=1}^n {\Big (} C_{hk}^i + T_{[hk]}^i(x) {\Big )}
x_h x_k     -    x_i  \sum_{k=1}^n x_k \pb
\label{evolution eq eta = 1}
\end{equation}

In order to design within the general framework 
$(\ref{evolution eq eta = 1})$
a specific model (or a specific family of models), we need to further characterize the expressions 
of the direct transition probability densities $C_{hk}^i$ 
and 
the indirect transition variation densities $T_{[hk]}^i(x)$. A conceivable choice is given next.

\smallskip

We represent the direct transition probability densities $C_{hk}^i$ as
$$
C_{hk}^i = a_{hk}^i + b_{hk}^i \vb
$$
where
the term $a_{hk}^i$ 
expresses the probability density that an $h$-individual
will belong to the $i$-th class
after an encounter with a $k$-individual,
when such an encounter does not produce any change
of class
and the term $b_{hk}^i$ expresses the 
density variation in the $i$-th class
of an $h$-individual interacting with a $k$-individual.

Accordingly, the only nonzero elements
$a_{hk}^i$ are
$$
a_{ij}^i = 1 \pb
$$

To define the elements $b_{hk}^i$,
we introduce the matrix $P$, whose elements $p_{h,k}$ 
express the probability that  in an encounter between an $h$-individual and a $k$-individual,
the one who pays is the $h$-individual. In view of the possibility that
to some extent the two individuals do not really interact, the $p_{h,k} $ are required to satisfy $0 \le p_{h,k} \le 1$ and, 
of course, $p_{h,k} + p_{k,h} \le 1$. 
Apart from that, there is a certain arbitrariness in the construction of the matrix $P$.
The only additional requirement is
to put each element in the first row, as well as each element
but the very last one in the last column, equal to zero.
The reason for that is the non existence of a class lower than the first
nor a class higher than the $n$-th: we cannot admit the possibility 
for $1$-individuals [respectively, for $n$-individuals] to move back to
a lower class [respectively, to advance, passing in a higher class]. Observing that 
in the presence of interactions between two individuals of the same class, the average wealth of the two 
remains the same if we only keep into account the direct transition, and may only change
because of the taxation and redistribution contributions,
we assume that individuals of class $1$ never pay.
We also assume that individuals of class $n$ 
never receive money

An encounter between an $h$-individual and a $k$-in\-di\-vid\-u\-al where $h \ne k$, with 
$h \ge 2$ and $k \le n-1$
and
the
$h$-individual paying, 
produces 
the elements\footnote{\ Since the classes 
are characterized by an average income, the terms $b_{hh}^i$ are equal to zero
for any $h$ and $i$.} 
\begin{eqnarray}
& & b_{hk}^{h-1} = p_{h,k} \, S \, \frac{1-\tau_k}{r_h - r_{h-1}} \vb  \quad
b_{hk}^h = - p_{h,k} \, S \, \frac{1-\tau_k}{r_h - r_{h-1}} \vb \nonumber \\
& & b_{kh}^{k+1} = p_{h,k} \, S \, \frac{1-\tau_k}{r_{k+1} - r_{k}} \vb \quad
b_{kh}^k  = - p_{h,k} \, S \, \frac{1-\tau_k}{r_{k+1} - r_{k}} \pb \nonumber
\label{bhk}
\end{eqnarray} 
Hence, the possibly nonzero elements
$b_{hk}^i$ are of the form 
\begin{eqnarray}
& & b_{i+1,k}^{i} = p_{i+1,k} \, S  \, \frac{1-\tau_k}{r_{i+1} - r_{i}} \vb \nonumber \\
& & b_{i,k}^i = - \, p_{k,i} \, S \, \frac{1-\tau_i}{r_{i+1} - r_{i}} 
    - \, p_{i,k} \, S \, \frac{1-\tau_k}{r_{i} - r_{i-1}} \vb \nonumber \\
& & b_{i-1,k}^i =p_{k,i-1} \, S \, \frac{1-\tau_{i-1}}{r_{i} - r_{i-1}} \vb
\label{b}
\end{eqnarray} 
where 
the expression for $b_{i+1,k}^{i}$ in $(\ref{b})$ holds true
for $i+1 \ne k$, and
for $i \le n-1$ and $k\le n-1$;
in the expression for $b_{i,k}^i$, which is only present whenever $i \ne k$,
the first addendum is effectively present only 
provided $i \le n-1$ and $k \ge 2$ and
the second addendum
only 
provided $i \ge 2$ and $k \le n-1$, 
and the expression for $b_{i-1,k}^i$ holds true for
for $i-1 \ne k$, and
for $i \ge 2$ and $k\ge 2$.

\smallskip

We express the indirect transition variation densities $T_{[hk]}^i(x)$ as
$$
T_{[hk]}^i(x) =  
U_{[hk]}^i(x) + V_{[hk]}^i(x)
\vb
$$
where 
\begin{equation}
U_{[hk]}^i(x) =  
\frac{p_{h,k} \, S \, \tau_k}{\sum_{j=1}^{n} x_{j}} {\bigg (}  \frac{x_{i-1}}{r_i - r_{i-1}} -   \frac{x_{i}}{r_{i+1} - r_{i}} {\bigg )}
\label{U_{[hk]}^i(x)}
\end{equation}
represents
the variation density corresponding to the advancement
from a class to the subsequent one, due to the benefit of taxation
and
\begin{equation}
V_{[hk]}^i(x)
=  p_{h,k} \, S \, \tau_k \, 
{\bigg (} 
\frac{\delta_{h,i+1}}{r_h - r_{i}} \, - \, \frac{\delta_{h,i}}{r_h - r_{i-1}}
{\bigg )} 
\, \frac{\sum_{j=1}^{n-1} x_{j}}{\sum_{j=1}^{n} x_{j}}
\vb
\label{V_{[hk]}^i(x)}
\end{equation}
with
$\delta_{h,k}$ denoting the 
{\slshape Kronecker delta},
accounts for 
the variation density corresponding to the retrocession
from a class to the preceding one, due to the payment
of some tax.
In the r.h.s. of 
$(\ref{U_{[hk]}^i(x)})$ and $(\ref{V_{[hk]}^i(x)})$,
$h >1$ 
and
the terms 
involving the index $i-1$ [respectively, $i+1$]
are effectively present only provided $i-1 \ge 1$ 
[respectively, $i+1 \le n$].  

Notice that for technical reasons, in the model under consideration,
the effective amount of money paid as tax
relative to an exchange of $S (1 - \tau_k)$
between two individuals
and then redistributed among classes
is given by $S \, \tau_k \,({\sum_{j=1}^{n-1} x_{j}})/{(\sum_{j=1}^{n} x_{j}})$
instead of $S \, \tau_k$.

\smallskip

A general theorem proved in \cite{Ber} ensures
that,
with the present choice of parameters, in correspondence to any initial condition
$x_0 = (x_{01} , \ldots , x_{0n})$, 
for which $x_{0i} \ge 0$ and $\sum_{i=0}^n x_{0i} = 1$,
a unique solution $x(t) = (x_1(t),\ldots,x_n(t))$ of $(\ref{evolution eq eta = 1})$ exists,
which is defined for all $t \in [0,+\infty)$, satisfies $x(0) = x_0$ and also
\begin{equation}
x_{i}(t) \ge 0 \ \hbox{for} \ i \in \Gamma_n \ \hbox{and} \ \sum_{i=0}^n x_{i}(t) = 1 \ \hbox{for all} \ t \ge 0 \pb 
\label{solution in the future}
\end{equation}
With reference to this well-posedness result, we observe that in this context the only
meaningful
initial data, and solutions as well, are those with non negative components.
Also, the constraint 
$\sum_{i=0}^n x_{i0} = 1$ expresses nothing but a normalization.
Hence, we conclude by $(\ref{solution in the future})$ that
the 
solutions of interest
are in fact 
distribution functions.
Furthermore, we emphasize that in view of
$(\ref{solution in the future})$,
the expressions of the $U_{[hk]}^i(x)$ and $V_{[hk]}^i(x)$ in 
$(\ref{U_{[hk]}^i(x)})$
and 
$(\ref{V_{[hk]}^i(x)})$
can be simplified. 

We may from now on 
consider, instead of $(\ref{evolution eq eta = 1})$, the system of differential equations
\begin{equation}
{{d x_i} \over {d t}} =  
\sum_{h=1}^n \sum_{k=1}^n {\Big (} C_{hk}^i + T_{[hk]}^i(x) {\Big )}
x_h x_k     -    x_i  
\vb
\quad i \in {\Gamma}_n \vb
\label{simplified evolution eq eta = 1}
\end{equation}
where the terms $T_{[hk]}^i(x)$ are linear in the variables $x_j$.
The equations in $(\ref{simplified evolution eq eta = 1})$ 
have a polynomial right hand side,
containing cubic terms as the highest degree ones.
 
Due to the fact that the value of $n$ and the parameters $r_k, \tau_k, p_{h,k}$ 
are still to be fixed,
the equations $(\ref{simplified evolution eq eta = 1})$ 
actually describe a family of models 
rather than a single model.

%%%%%%%%% %%%%%%%%% %%%%%%%%% %%%%%%%%% %%%%%%%%%
\section{Different choices of the parameters}
\label{diff cho}
%%%%%%%%% %%%%%%%%% %%%%%%%%% %%%%%%%%% %%%%%%%%%

We now exploit the 
flexibility which characterizes the equations $(\ref{simplified evolution eq eta = 1})$ 
and take into consideration three different possible choices for the values of the 
parameters $p_{h,k}$.
Our aim is to try and see what is the effect of such different choices on the shape of the 
large wealth part of the asymptotic income distribution.

%%%
\subsection{Three different cases}
%%%

In each of the three considered cases 
we take the elements of  the matrix $P$ lying 
on the main diagonal, those on the first row and those on the $n$-th column to be equal to zero:
$$
\begin{array}{llll}
p_{j,j} = 0 \quad & \hbox{for} \ j \in \{1, ..., n\} \vb \nonumber \\
p_{1,k} = 0 \quad & \hbox{for} \ k \in \{1, ..., n\} \vb \nonumber \\
p_{h,n} = 0 \quad & \hbox{for} \ h \in \{1, ..., n\} \pb \nonumber 
\label{P(common)}
\end{array} 
$$
Apart from that, we define the $p_{h,k}$

\bigskip

\noindent 
$\bullet$ \ \ in a first case, 

which we'll call the {\it{Case FSP}}, 

({\it{FSP}} for {\it{fixed saving propensity}}),

as
$$
p_{h,k} =({1}/{4}) \, {r_h}/{r_n}
$$

with the exception of the terms
$$
\begin{array}{llll}
p_{h,1} = ({1}/{2}) \, {r_h}/{r_n} \quad & \hbox{for} \ h \in \{2, ..., n\} \vb \nonumber \\
p_{n,k} = 
{1}/{2} \quad & \hbox{for} \ k \in \{1, ..., n-1\} \, ; \nonumber 
\label{P(I)}
\end{array} 
$$

\medskip 

\noindent 
$\bullet$ \ \ in a second case, 

which we'll call the {\it{Case TID}}, 

({\it{TID}} for {\it{trading individuals dependence}}),

as
$$
p_{h,k} = ({1}/{4}) \, \min \{r_h,r_k\}/{r_n}
$$

with the exception of the terms
$$
\begin{array}{llll}
p_{h,1} = 
({1}/{2}) \, {r_1}/{r_n}\quad & \hbox{for} \ h \in \{2, ..., n\} \vb \nonumber \\
p_{n,k} = 
({1}/{2})\, {r_k}/{r_n}\quad & \hbox{for} \ k \in \{1, ..., n-1\} \, ; \nonumber 
\label{P(II)}
\end{array} 
$$

\medskip

\noindent 
$\bullet$ \ \ in a third case, 

which we'll call the {\it{Case FAE}}, 

({\it{FAE}} for {\it{fixed amount exchange}}),

as
$$
p_{h,k} = {1}/{4}
$$

with the exception of the terms
$$
\begin{array}{llll}
p_{h,1} = {1}/{2} \quad & \hbox{for} \ h \in \{2, ..., n\} \vb \nonumber \\
p_{n,k} = {1}/{2} \quad & \hbox{for} \ k \in \{1, ..., n-1\} \, ; \nonumber 
\label{P(III)}
\end{array} 
$$

\bigskip

Choosing the $p_{h,k}$ as in the {\it{Case FSP}}
amounts to assume, in view of $(\ref{b})$, 
that each individual (apart from those belonging to the poorest and the richest class, for which particular rules hold true) 
has the same saving pro\-pen\-sity. Indeed, the frequency of interactions is uniform
and
the effect of the mechanism described above is 
the same as if in any trade each individual would pay a small quantity
proportional to her income,
the proportion rate being the same for everybody.
Such a homogeneity in spending and hence in saving is replaced by heterogeneity both in the  {\it{Cases}} {\it{TID}} and {\it{FAE}}.
In particular, the {\it{Case TID}} expresses the conjecture that,
%typically
when two individuals trade, the exchanged amount of money is proportional to the wealth of the poorest of the two.

\smallskip

The
average incomes $r_j$
and the taxations rates $\tau_j$ are taken here 
to be given by
$r_j = 10 j$ and 
\begin{equation}
\tau_j = \tau_{min} +  \frac{j - 1}{n-1} \, (\tau_{max} - \tau_{min}) 
\label{progressivetaxrates}
\end{equation}
for $j \in \Gamma_n$,
where $ \tau_{min} = 10/100$ and $\tau_{max} = 45/100$.

\medskip

According to the result of
a large number of simulations 
(see also \cite{Ber} and \cite{BerMod}),
once the value of $n$ and the parameters $r_j$ and $\tau_j$
are chosen,
in correspondence to any fixed value $\mu \in [r_1,r_n]$ of the global wealth
a stationary distribution exists, which 
is the asymptotic trend of all solutions of $(\ref{simplified evolution eq eta = 1})$
with initial conditions $x_0 = (x_{01} , \ldots , x_{0n})$ 
satisfying
$x_{0i} \ge 0$, $\sum_{i=0}^n x_{0i} = 1$
and
$\sum_{i=1}^n r_i x_{0i} = \mu$.

\medskip

The influence on such distribution of the different interaction rules relative to the {\it{Cases}} {{\it{FSP}}, {\it{TID}} and {\it{FAE}}} is well illustrated by 
the Figures $1, 2$,
referring to the case $n=25$, 
just a sample among several similar ones. 
In each of these figures 
two panels relative to the {\it{Case FSP}} in the first row,
two relative 
to the  {\it{Case TID}} in the second row,
and
two relative 
to the {\it{Case FAE}} in the third row
are reported.
The panels on the left show the histograms corresponding to 
an initial condition; the panels on the right show the corresponding asymptotic stationary distribution.
We point out here that the histograms are scaled differently from panel to panel.
The initial conditions are chosen
randomly within distributions for which
lower income classes contain a high percentage of individuals, which seems to be a rather realistic assumption.
In each of the two figures they are taken to be the same for the three simulations 
(each one referring to a different model) precisely to emphasize the 
role played by the different parameters in the formation of the shape of the asymptotic distribution.

%%%%%%%%%%%%%%%%%%%%%%%%%%%
\begin{figure*}
  \begin{center}
  \includegraphics[width=4cm,height=2cm]  {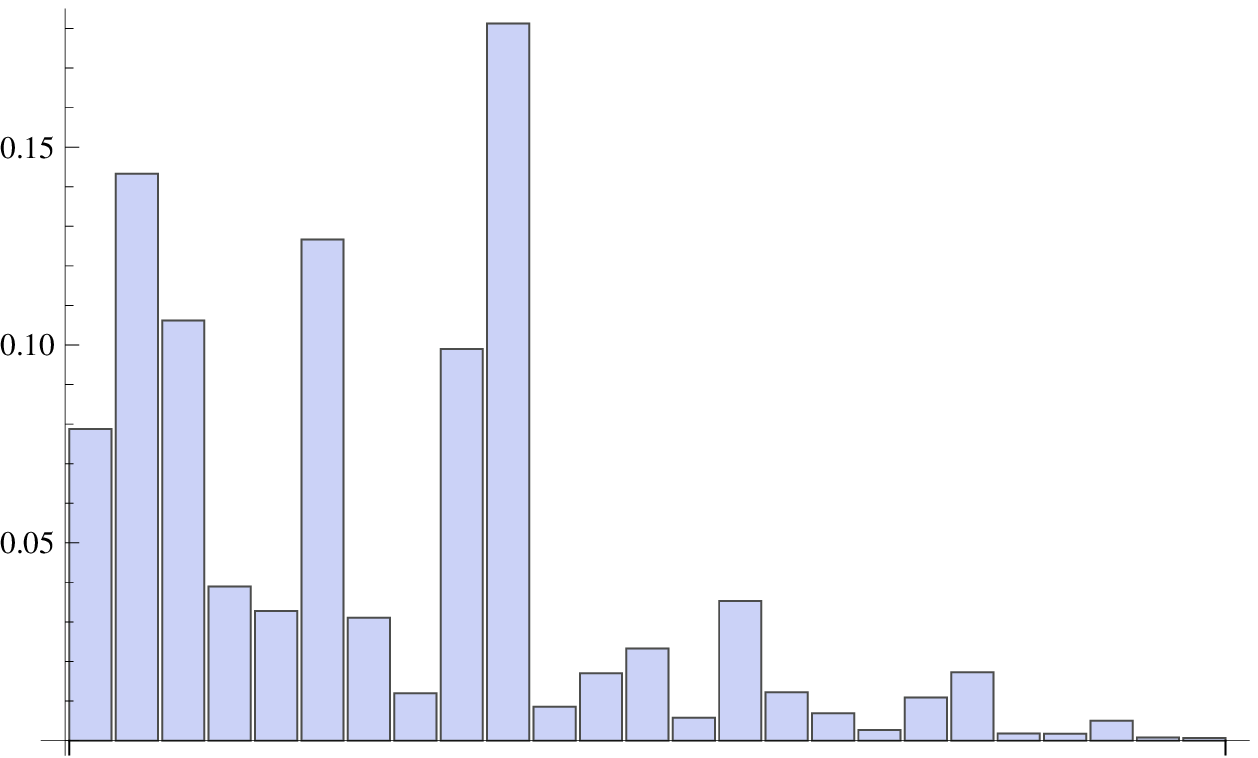}
  \hskip0.5cm
  \includegraphics[width=4cm,height=2cm]  {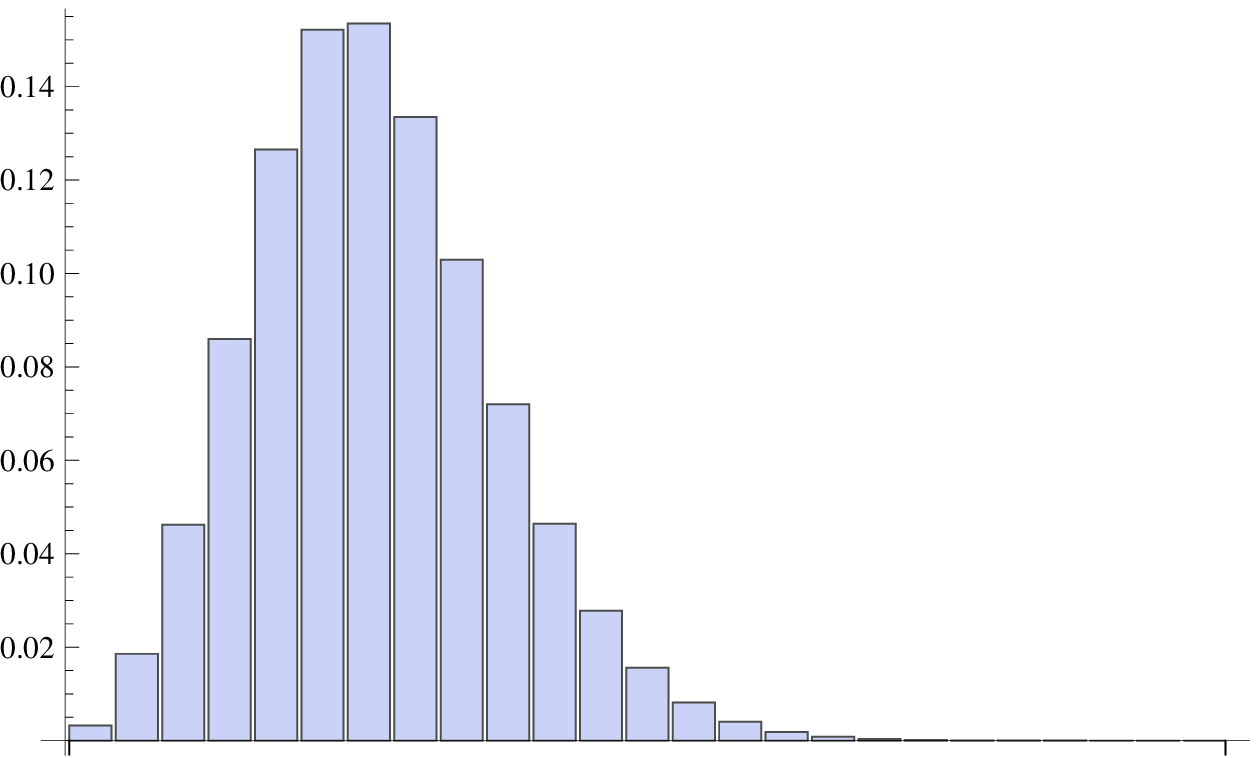}
  \end{center}
  \begin{center}
  \includegraphics[width=4cm,height=2cm]  {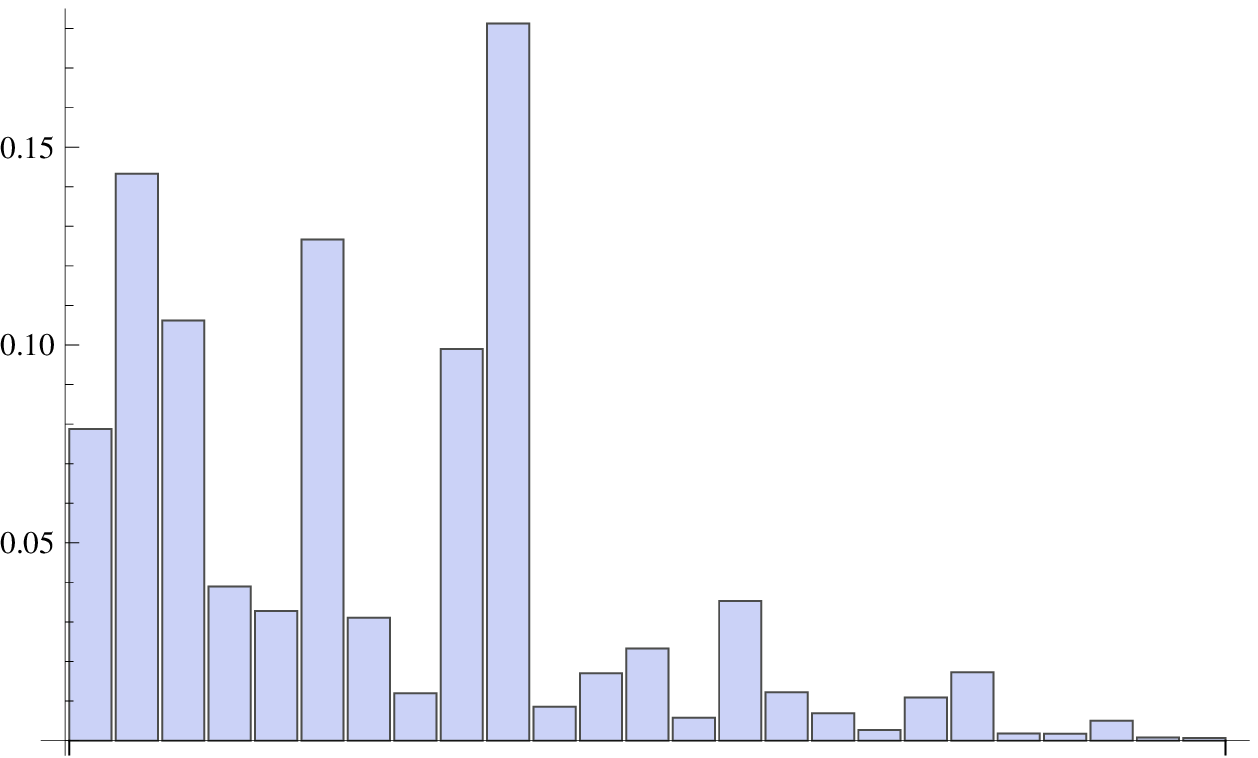}
  \hskip0.5cm
  \includegraphics[width=4cm,height=2cm]  {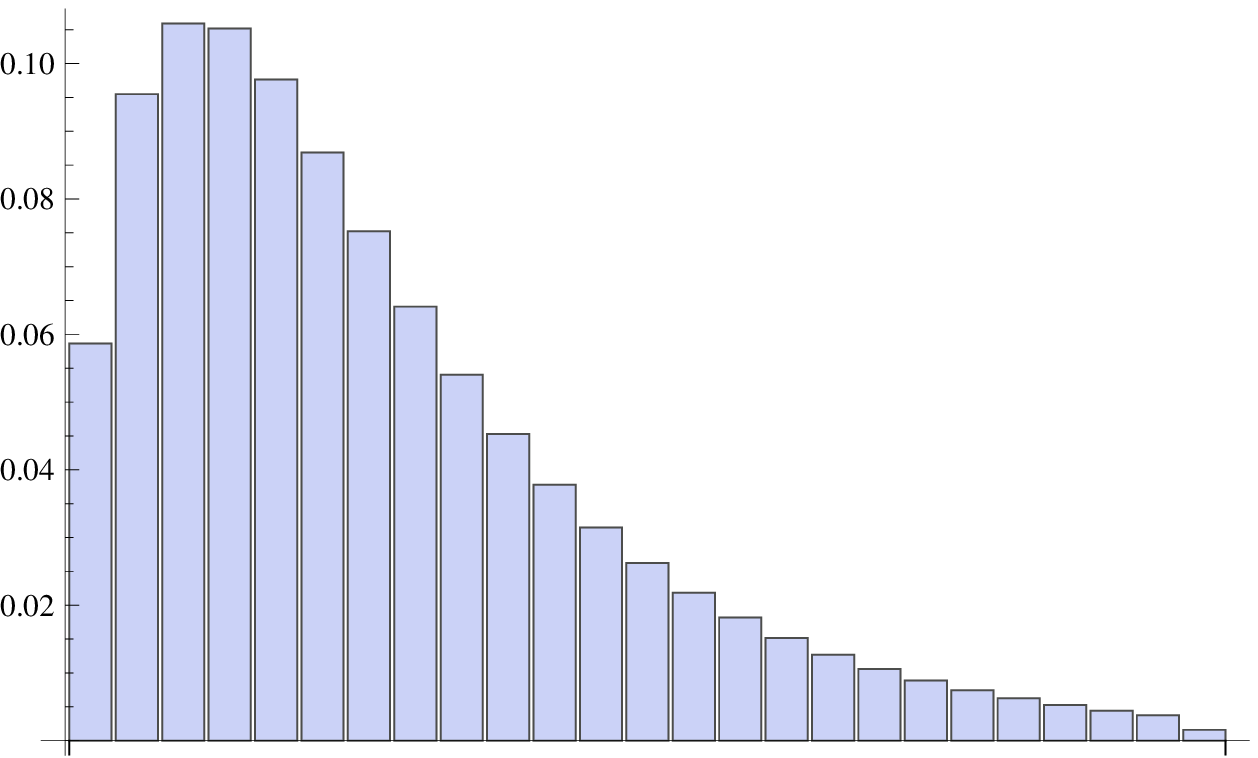}
  \end{center}
   \begin{center}
  \includegraphics[width=4cm,height=2cm]  {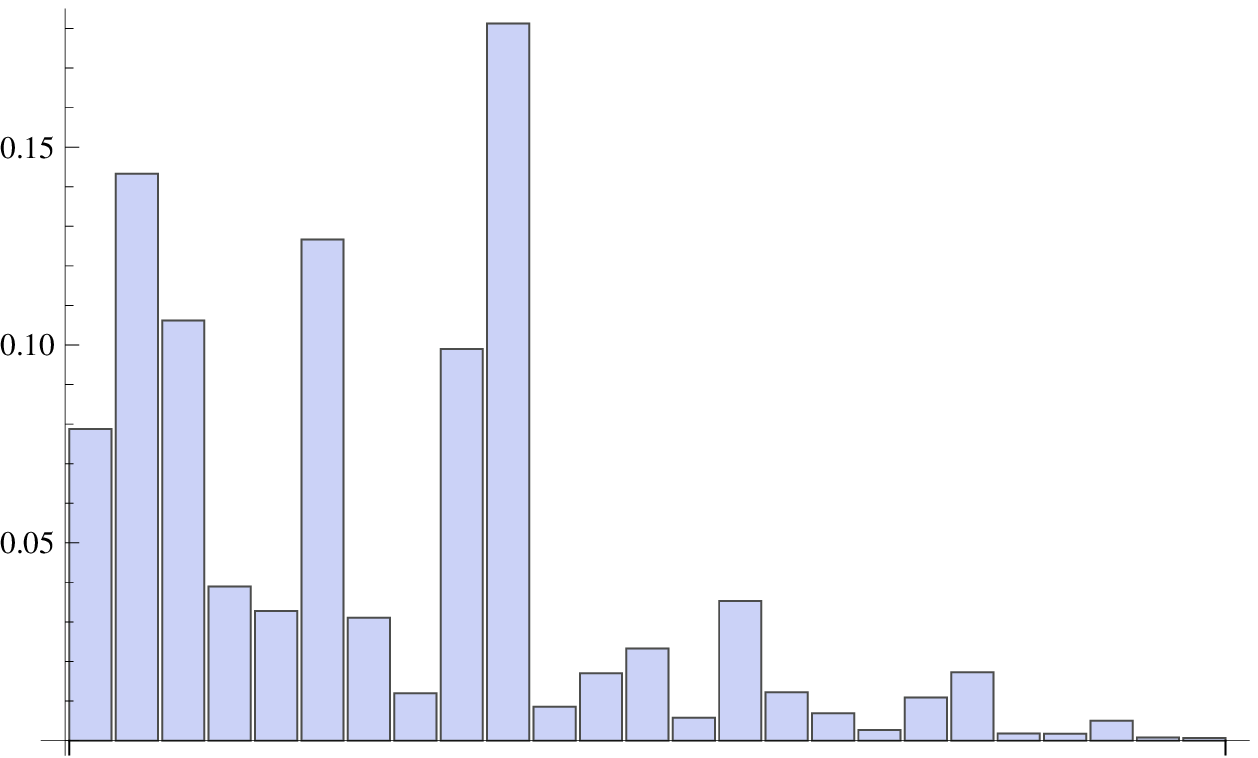}
  \hskip0.5cm
  \includegraphics[width=4cm,height=2cm]  {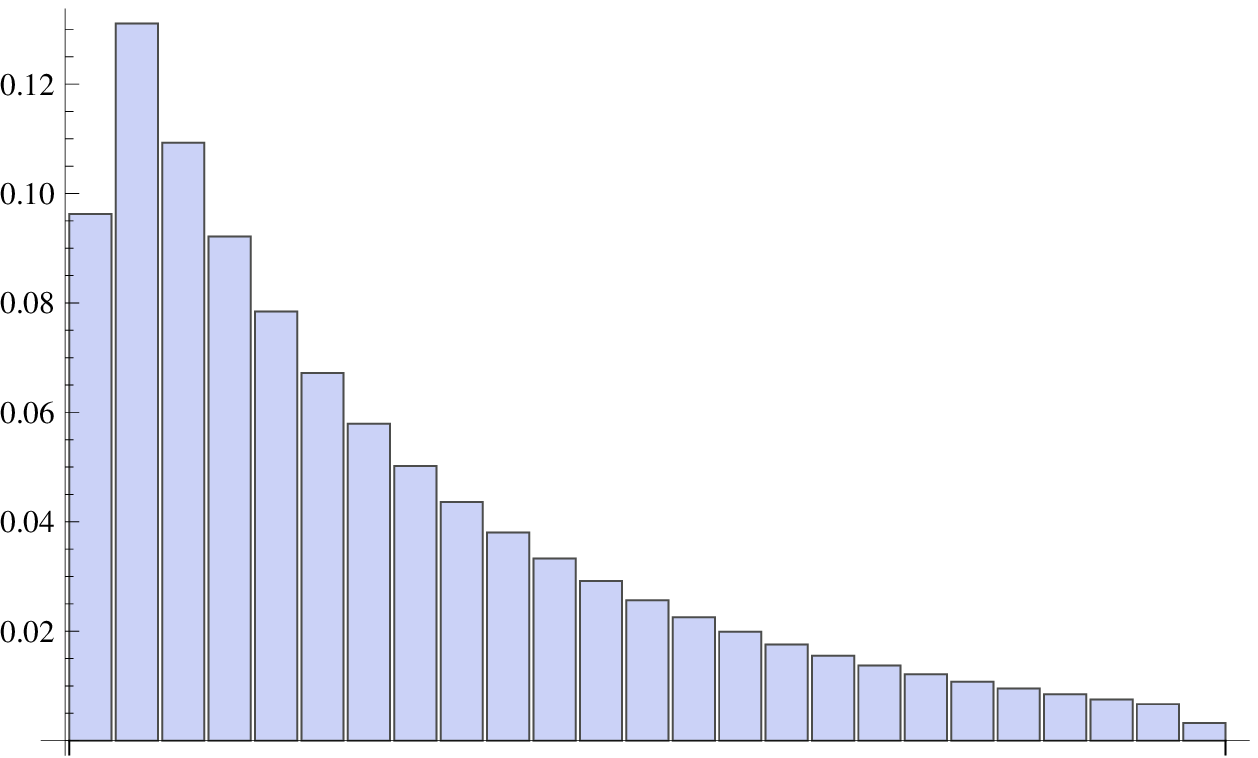}
  \end{center}
\caption{Initial (on the left) and long-time equilibrium (on the right) distributions for the models at issue:
the first row refers to the {\it{Case FSP}} ({\it{fixed saving propensity}}),
the second row to the  {\it{Case TID}} ({\it{trading individuals dependence}}),
the third row to the {\it{Case FAE}} ({\it{fixed amount exchange}}).
Notice that
the histograms are scaled differently on different pictures.} 
\end{figure*}
%%%%%%%%%%%%%%%%%%%%%%%%%%%

%%%%%%%%%%%%%%%%%%%%%%%%%%%
\begin{figure*}
  \begin{center}
  \includegraphics[width=4cm,height=2cm]  {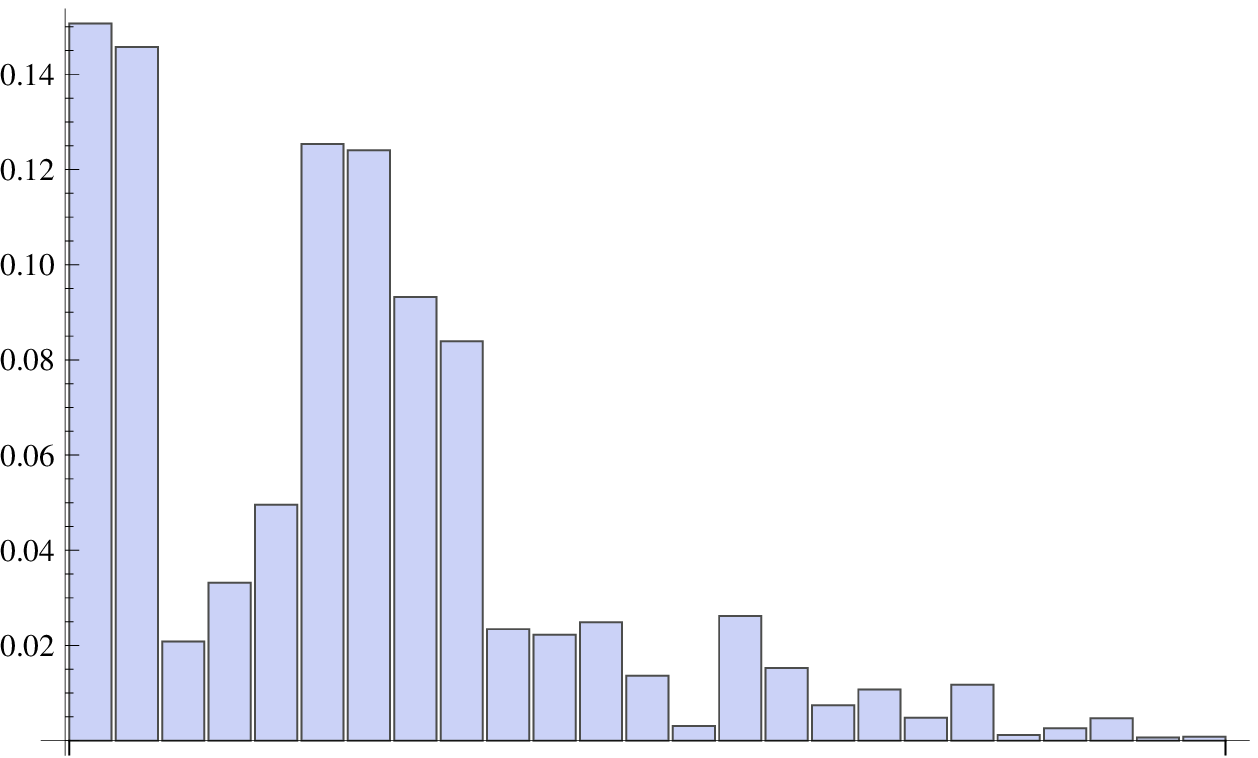}
  \hskip0.5cm
  \includegraphics[width=4cm,height=2cm]  {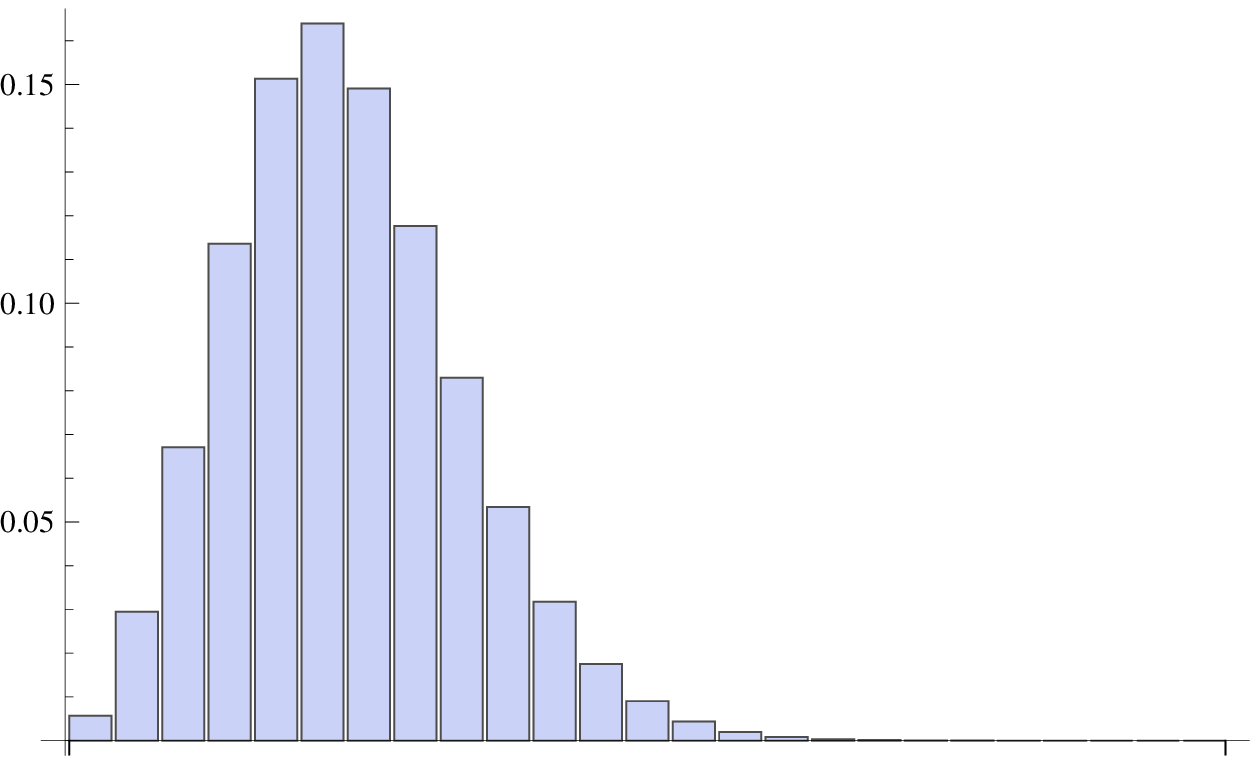}
  \end{center}
  \begin{center}
  \includegraphics[width=4cm,height=2cm]  {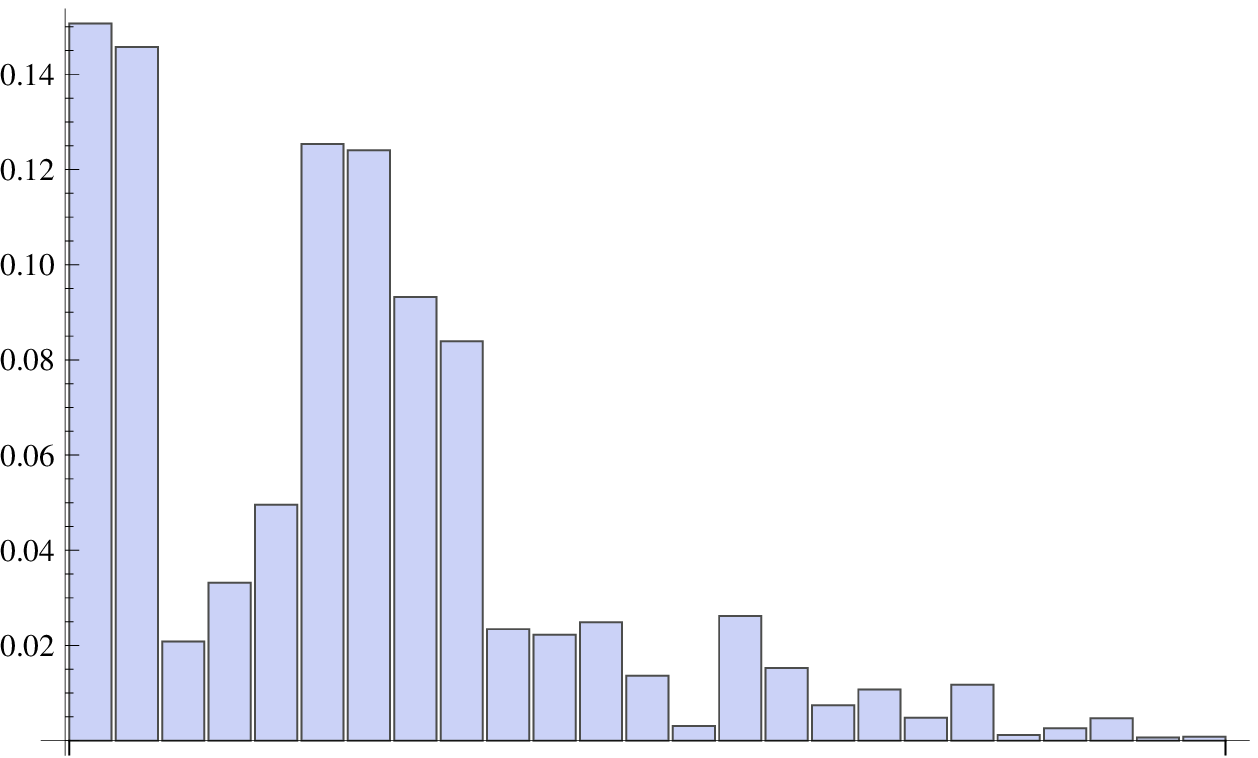}
  \hskip0.5cm
  \includegraphics[width=4cm,height=2cm]  {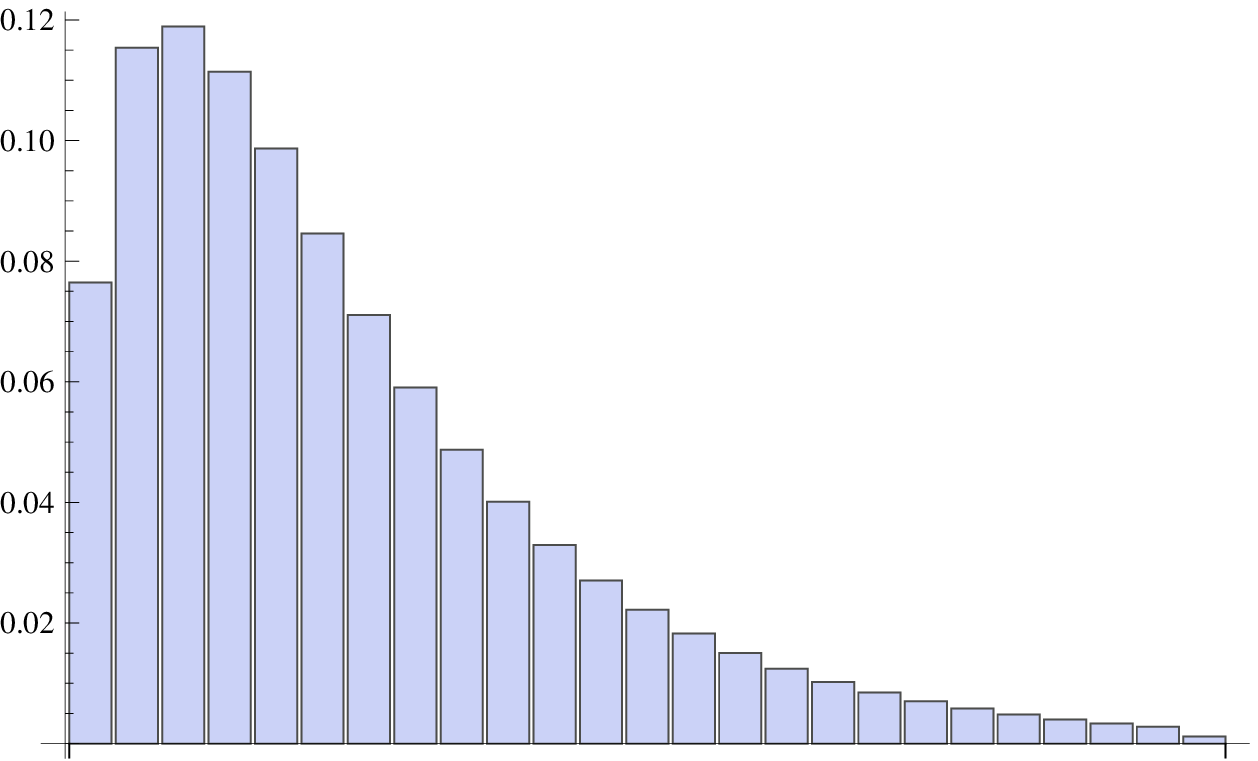}
  \end{center}
   \begin{center}
  \includegraphics[width=4cm,height=2cm]  {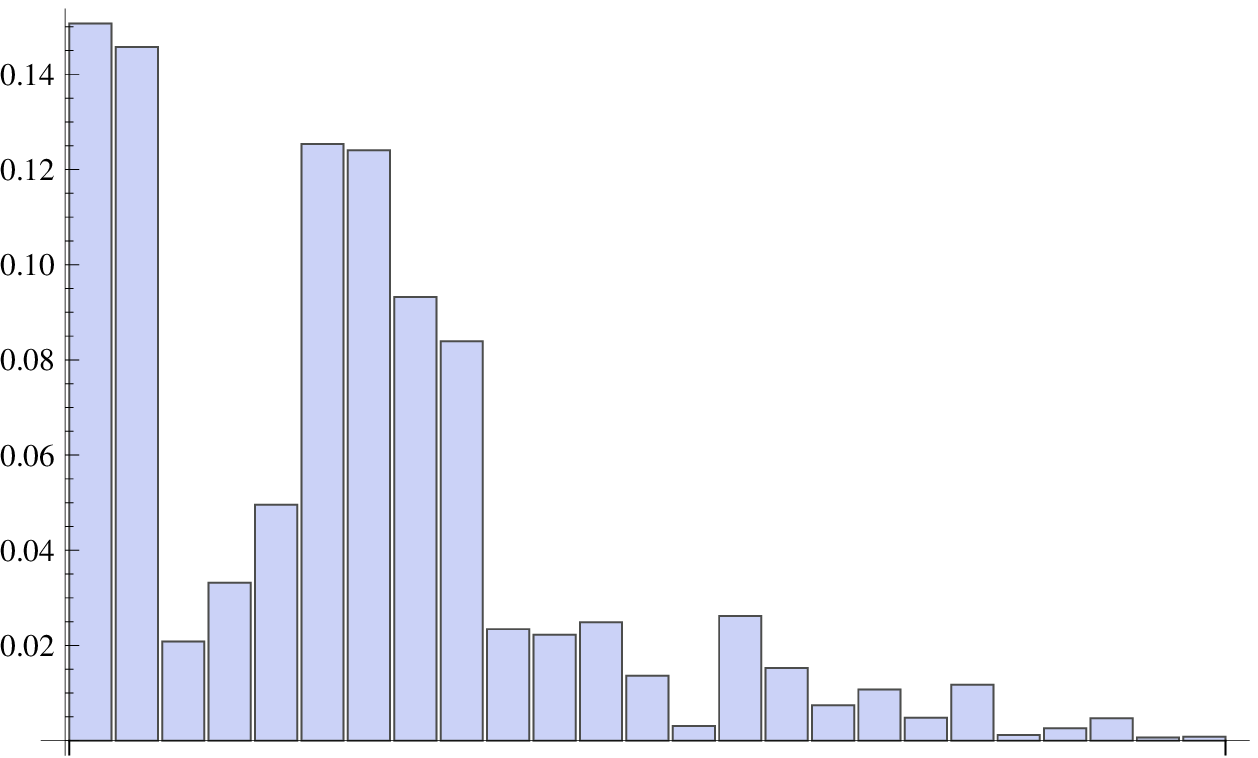}
  \hskip0.5cm
  \includegraphics[width=4cm,height=2cm]  {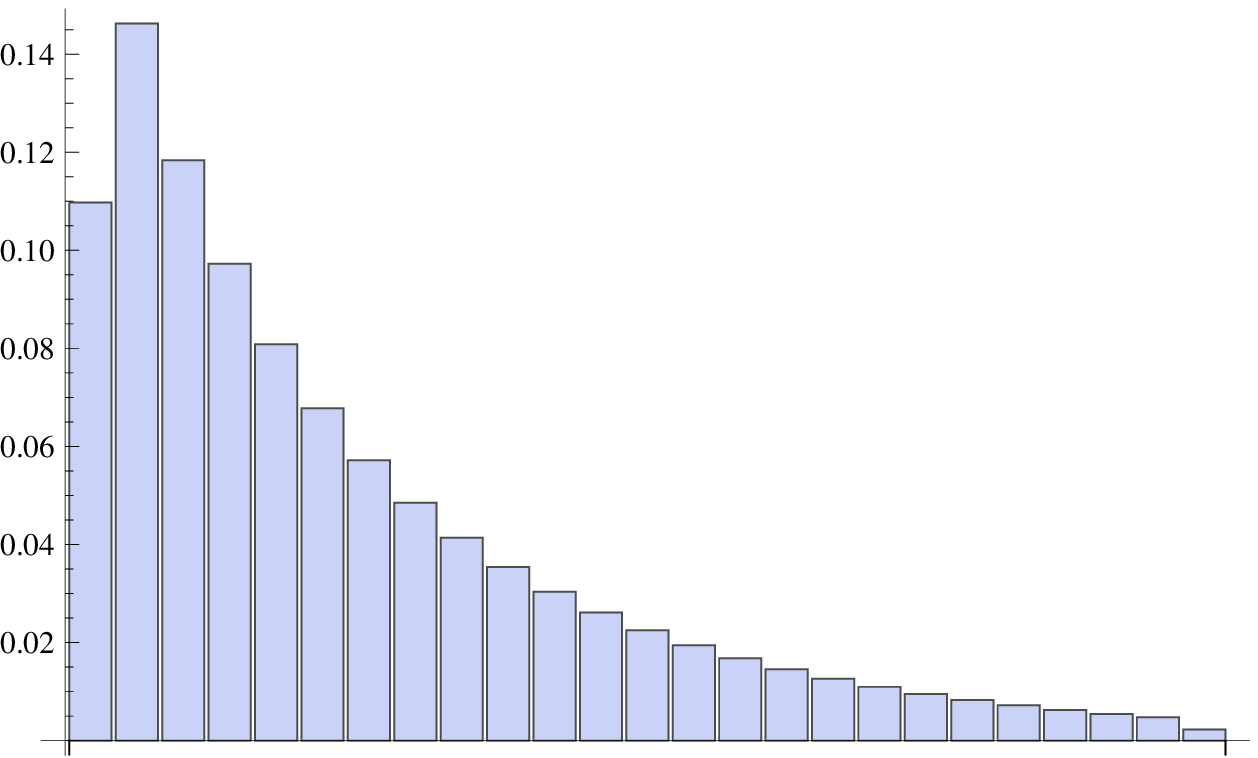}
  \end{center}
\caption{Initial (on the left) and long-time equilibrium (on the right) distributions for the models at issue:
the first row refers to the {\it{Case FSP}} ({\it{fixed saving propensity}}),
the second row to the  {\it{Case TID}} ({\it{trading individuals dependence}}),
the third row to the {\it{Case FAE}} ({\it{fixed amount exchange}}).
Notice that
the histograms are scaled differently on different pictures.} 
\end{figure*}
%%%%%%%%%%%%%%%%%%%%%%%%%%%

The Figures $1$ and $2$ suggest that a fat tail exists in the  {\it{Cases}} {\it{TID}} and {\it{FAE}}, and not in the {\it{Case FSP}}.
As a more,
a log-log plot of the right section of the asymptotic stationary distributions is in agreement with
a power law type decrease for the {\it{Cases}} {\it{TID}} and {\it{FAE}}, while this does not happen in the {\it{Case FSP}}.

To gain some accuracy, we tried several simulations in correspondence to a different choice of the average incomes: 
in particular,
we took them to be nonlinear: 
$r_j = 10 \, (1,67)^j$ for $j \in \Gamma_{25}$,
so as to have $r_{24}/r_{15} \sim100$.
The expectation was confirmed: no power law behavior is seen in the {\it{Case FSP}},
while a clear straight line appears in the log-log plots of the {\it{Cases}} {\it{TID}} and {\it{FAE}}.
As a sample of the result of several simulations, three log-log plots
relatives to the {\it{Cases}} {{\it{FSP}}, {\it{TID}} and {\it{FAE}}} are shown in Figure $3$.

%%%%%%%%%%%%%%%%%%%%%%%%%%%
\begin{figure*}
  \begin{center}
  \includegraphics[width=5.5cm,height=3cm]  {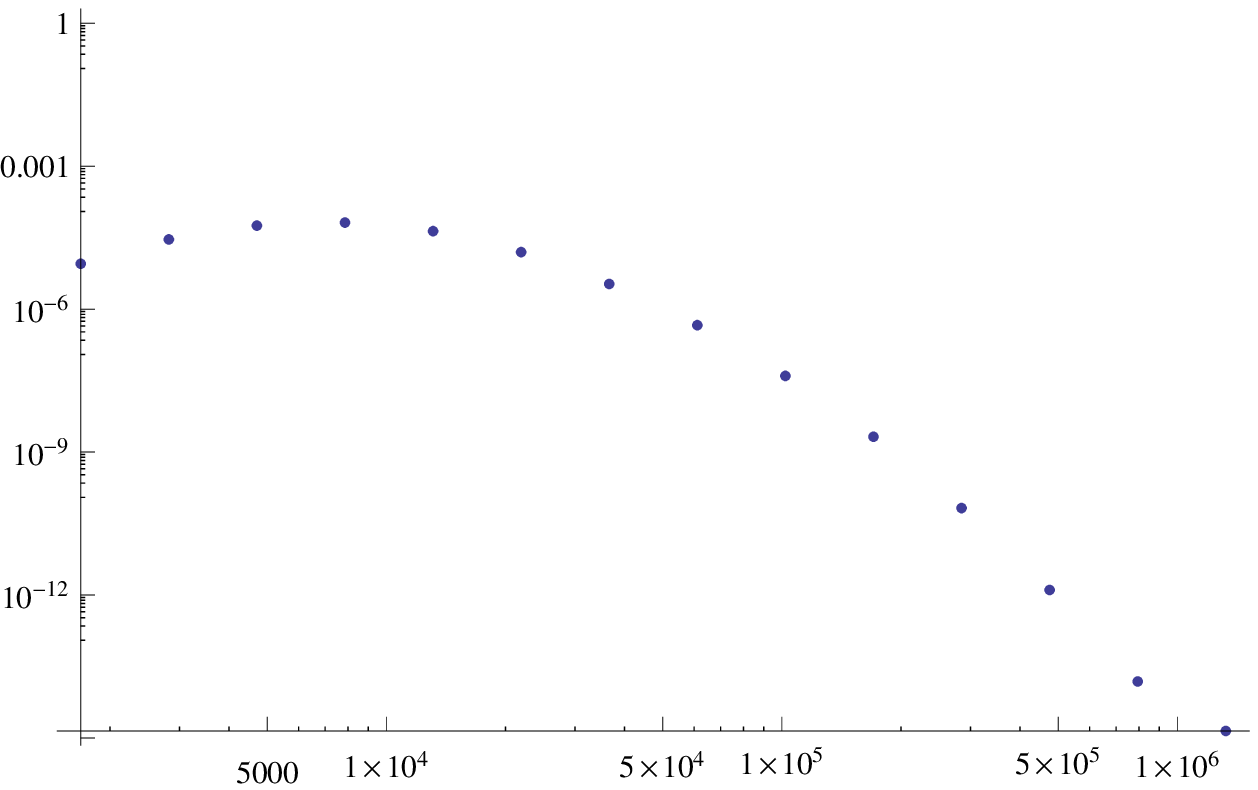}
  \hskip0.35cm
  \includegraphics[width=5.5cm,height=3cm]  {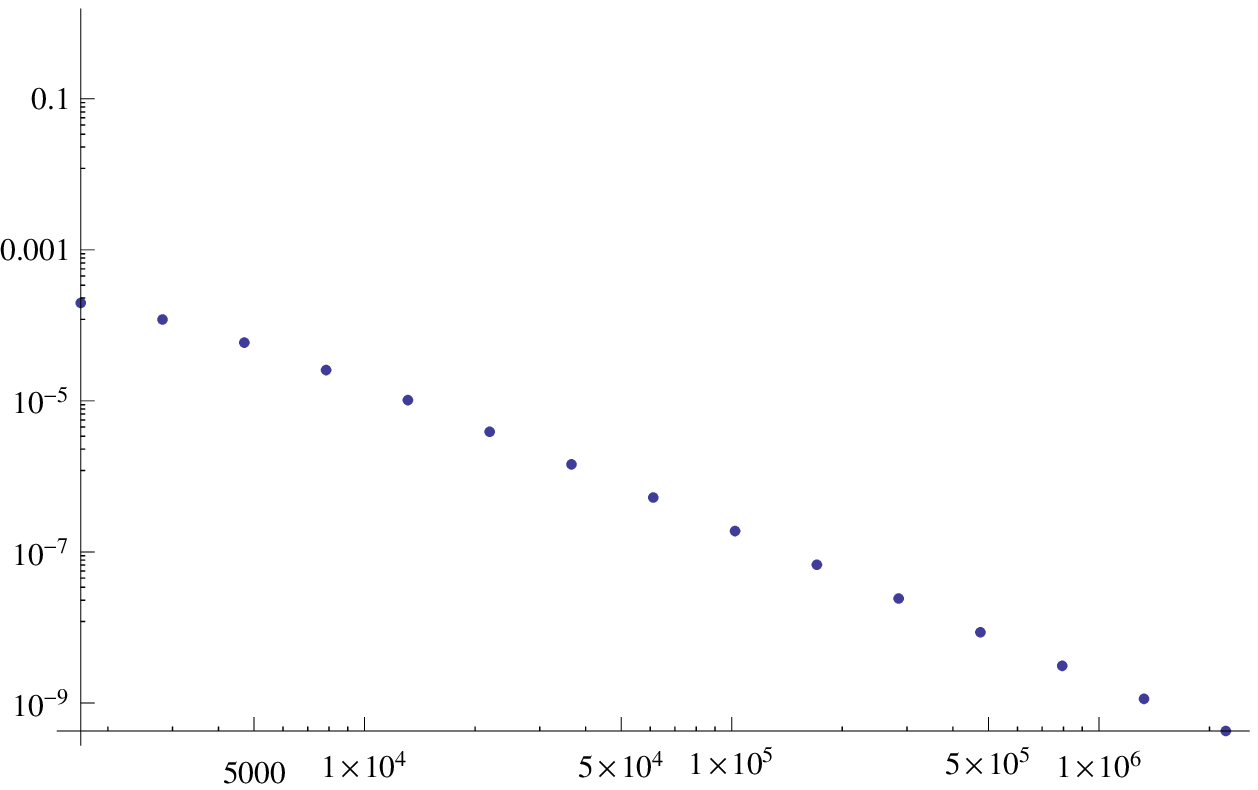}
   \hskip0.35cm
  \includegraphics[width=5.5cm,height=3cm]  {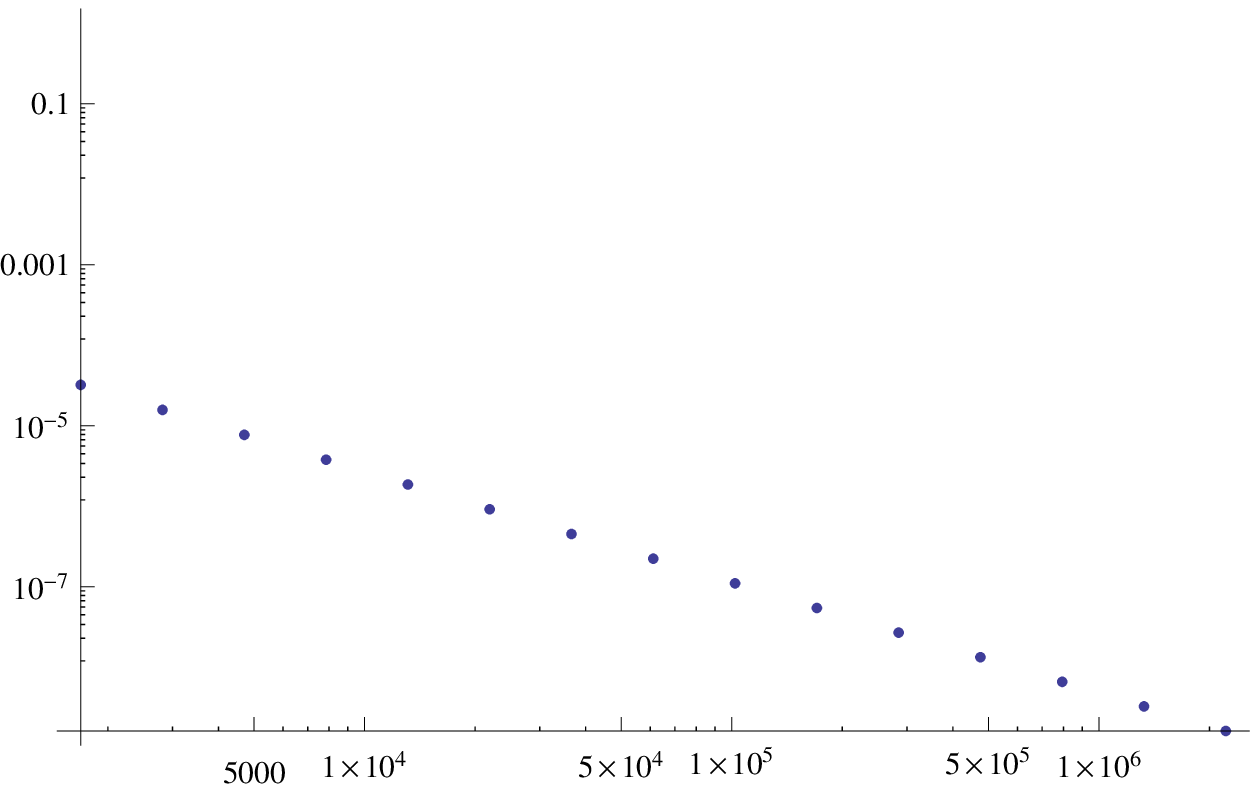}
  \end{center}
\caption{The (suitably rescaled) log-log plots of the right section of asymptotic stationary distributions
(in a case with average incomes $r_j = 10 \, (1,67)^j$)
in the three {\it{Cases}}:
{\it{FSP}},
{\it{TID}},
and {\it{FAE}}.
A clear straight line appears in the log-log plots of the {\it{Cases}} {\it{TID}} and {\it{FAE}}, while this 
does not happen in the {\it{Case FSP}}.} 
\end{figure*}
%%%%%%%%%%%%%%%%%%%%%%%%%%%

%%%
\subsection{Some remarks on finding power laws}
%%%

In general, simulations based on classes whose income
values increase linearly do not allow a precise determination
of the tail of the equilibrium distribution. It is well
known (see e.g. \cite{CleGal}) that in real statistical data a fraction 
of less than $2$-$3$ $\%$ of the population has an income distributed
according to a power law. But a power law tail can be clearly
distinguished from an exponential tail only over an income
range of two or three magnitude orders. This requires 
the introduction
in our model of class incomes growing more steeply than linearly; 
in this way the individuals of the $25$-th class
would not be,
for instance, $25$ times richer than those of the first class, but at least
$1000$ times richer.

In our previous work \cite{BerMod} we introduced for this purpose
class incomes which increase quadratically; in that case the
income of the $25$-th class was $25^2$ times larger than the income of the first class.
Notice that when the class income $r_i$ is not linear in the $i$-th class,
a rescaling of the discrete distribution function $x_i$ is
necessary, in order to preserve the correct relation between
the distribution density and the cumulative income distribution
\cite{BerMod}. It is possible to check that the cumulative distribution
$\sum_{i=1}^k x_i$ computed explicitly from the simulations tends
to coincide for large $k$ with the integral of the rescaled
density, provided one also takes into account certain slight
differences in the normalization (the discrete incomes are
normalized as $\sum_{i=1}^n x_i=1$, while the rescaled density
$f(r)$ is normalized as $\int_0^\infty f(r)dr=1$).

In the simulations for the investigation described here we employed, besides linear 
class incomes, incomes which increase exponentially as
$r_i=10 \, (1,67)^i$. The basis 1.67 is chosen in such a way as
to obtain an income increase by a factor 100 when the class
index $i$ increases by 10 (the tail of the distribution is
typically fitted to look for a power law in the range 
from $i=15$ to $i=25$). The rescaling factor to be applied to the
discrete distribution function $x_i$ is in this case proportional
to the reciprocal income $r_i^{-1}$; this means that a power law tail
maintains, after rescaling, the same power law form, with an
effective Pareto index equal to the index of the $x_i$ augmented
by $1$.

%%%
\subsection{A short survey of some wealth exchange models}
%%%
  
To put our analysis into an historical prespective,
we recall here that
several mathematical models of wealth exchange
processes display a Pareto tail in their equilibrium solutions.
In some cases the tail has only be observed numerically, in
others it is a part of an analytical solution. In some models
the Pareto index is a constant, in others it depends on some
parameters. Finally, some models are
conservative, the conserved quantity being the total wealth; 
in others the total wealth
increases in time. 

For instance, the model by Bouchaud and Mezard \cite{BouMez},
originally developed in physics for the description of polymers,
admits (when the exchange rate $J$ is the same for all agents)
an exact solution in the mean-field approximation.
The distribution function is in that case an Amo\-ro\-so curve
with a Pareto index $\alpha=1+J/\sigma^2$ which depends on the
exchange rate $J$ and on the variance $\sigma$ of a Gaussian
multiplicative process appearing in the model equations. The Gaussian
process simulates investment dynamics and is related to the
temporal change in the value of the stocks which, on the
average, increases in time.

The model by Scafetta, West and Picozzi \cite{ScaWesPic} is an evolution of
the model by Bouchaud and Mezard, which takes into account
several realistic details of the exchange process. Its
computer-generated wealth distribution functions can be
fitted by rational functions of the form $f(x)=ax^\gamma/(1+bx)^{\gamma+\delta}$,
where the Pareto index is $\alpha=\delta-1$.

In the model by Slanina \cite{Sla}, growing markets are
modeled by bringing in during each trading an extra amount of wealth,
which is proportional to the wealth of both agents participating
in the trading: this represents, realistically, the fact
that the rich can invest more, and with more return.
One finds in this case a power law distribution for the rich
of the form $f(x) \sim x^{-1.7}$.

In the classical conservative model by Chatterjee, Cha\-krabarti 
and Manna \cite{ChattChakrManna},
agents contribute only a fraction of their wealth for trading,
depending on their saving propensities, which differ among
agents. A Pareto tail is found, of the form $f(x) \sim x^{-2}$.

%\smallskip

Our model family some\-how re\-semb\-les that one by
Chatterjee, Cha\-krabarti and Manna \cite{ChattChakrManna}.
However, differently from it, it
allows to recover different Pareto indices (compare e.g. results in
\cite{BerMod} or see the data in the caption of the Figure $4$). 

%\smallskip

The question of taxation and redistribution was first studied in \cite{DraYak}, where it was modeled
through a Boltzmann equation. A different approach was developed in \cite{Gua}, 
encompassing two-step tradings, which consist of a wealth exchange analogous to an
inelastic binary \textquotedblleft collision \textquotedblright and of a redistribution 
of the lost wealth (the taxes) among the population.
The effect of the subsidies by the government on the equilibrium distribution is shown in both papers \cite{DraYak} and \cite{Gua} to cause a
shifting of the individuals from the lower income classes toward middle income classes. 
Accordingly, and differently with respect to what happens e.g. to Boltzmann-Gibbs distributions, 
the equilibrium distribution is seen to exhibit a maximum. The Figure $5$ of \cite{DraYak} and the Figure $2$ of \cite{Gua}
are qualitatively similar to the figures of this paper with the histograms representing the asymptotic equilibria.

%%%%%%%%% %%%%%%%%% %%%%%%%%% %%%%%%%%% %%%%%%%%%
\section{The $\kappa$ distribution as a reasonable fit}
\label{reasonable fit}
%%%%%%%%% %%%%%%%%% %%%%%%%%% %%%%%%%%% %%%%%%%%%

When the equilibrium income distribution $f(x)$ of a model is
found through numerical methods, it is customary to
characterize this function according to its general features
(e.g., monotonicity, presence of a maximum, value at zero) and to
compute some 
global statistical indices, as the variance, the Gini index, etc.. 
Finally, one tries to fit $f(x)$ with some known
multi-parameter distribution.

It is well known, for instance, that reversible kinetic
systems evolve towards a Gibbs function, independently from the
details of the transition probabilities, while non-reversible
systems with fixed saving propensity evolve towards
a Gamma function. 
For systems with a Pareto power law tail, 
four are, to our knowledge,
the fit functions which have been proposed in the literature:
the Amoroso distribution \cite{Amo}, 
the Pareto-Gompertz distribution \cite{ChaFigMouRib},
some rational functions of the form $f(x)=ax^\gamma/(1+bx)^{\gamma+\delta}$\cite{ScaWesPic},
and the $\kappa$-generalized distribution \cite{Kan}, \cite{CleGalKan}, \cite{CleDiMGalKan}.

%%%%%%%%%%%%%%%%%%%%%%%%%%%
\begin{figure*}
  \begin{center}
  \includegraphics[width=5.5cm,height=3cm]  {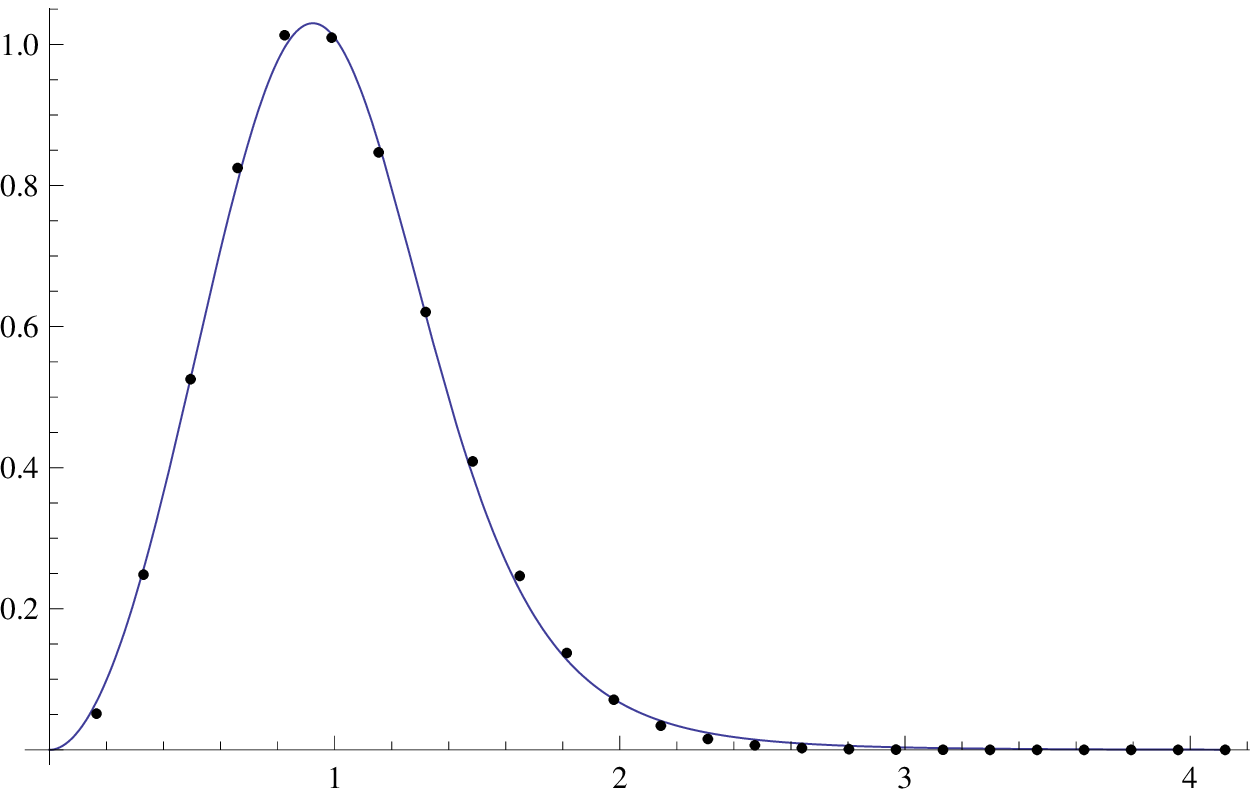}
  \hskip0.35cm
  \includegraphics[width=5.5cm,height=3cm]  {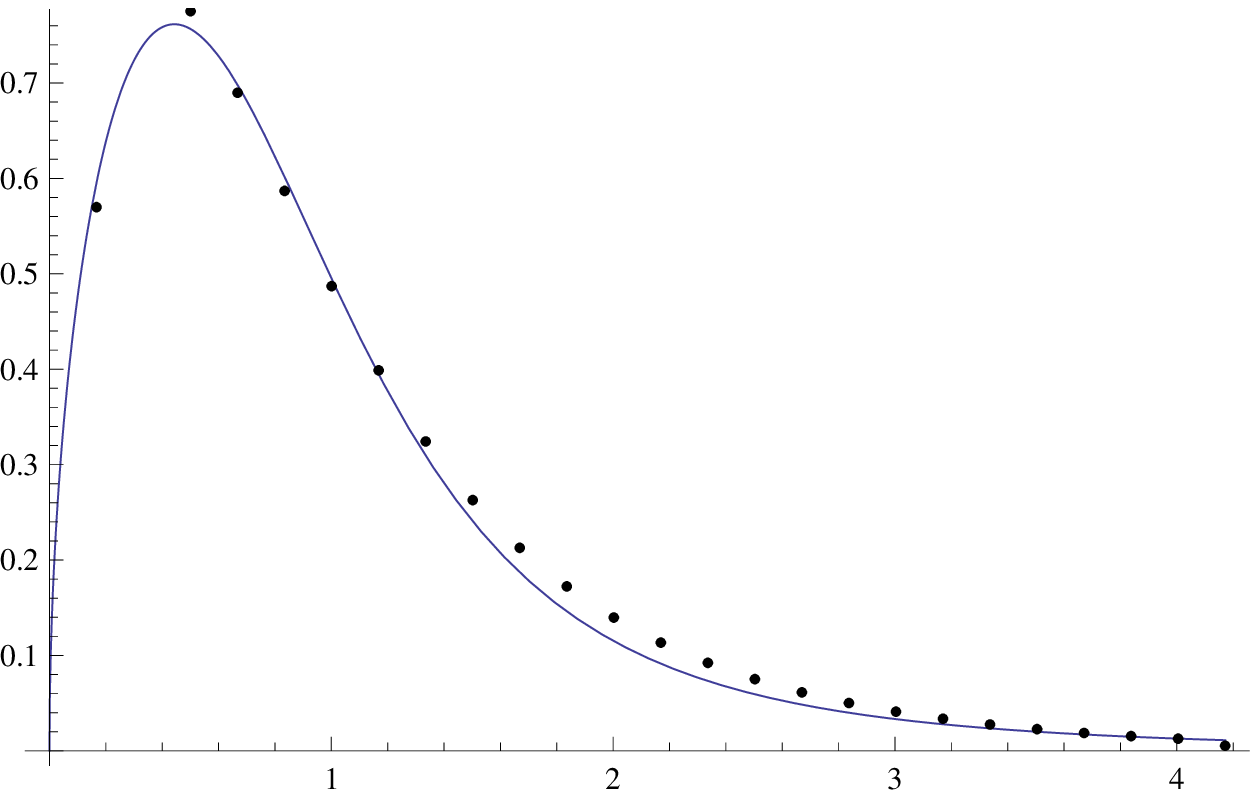}
   \hskip0.35cm
  \includegraphics[width=5.5cm,height=3cm]  {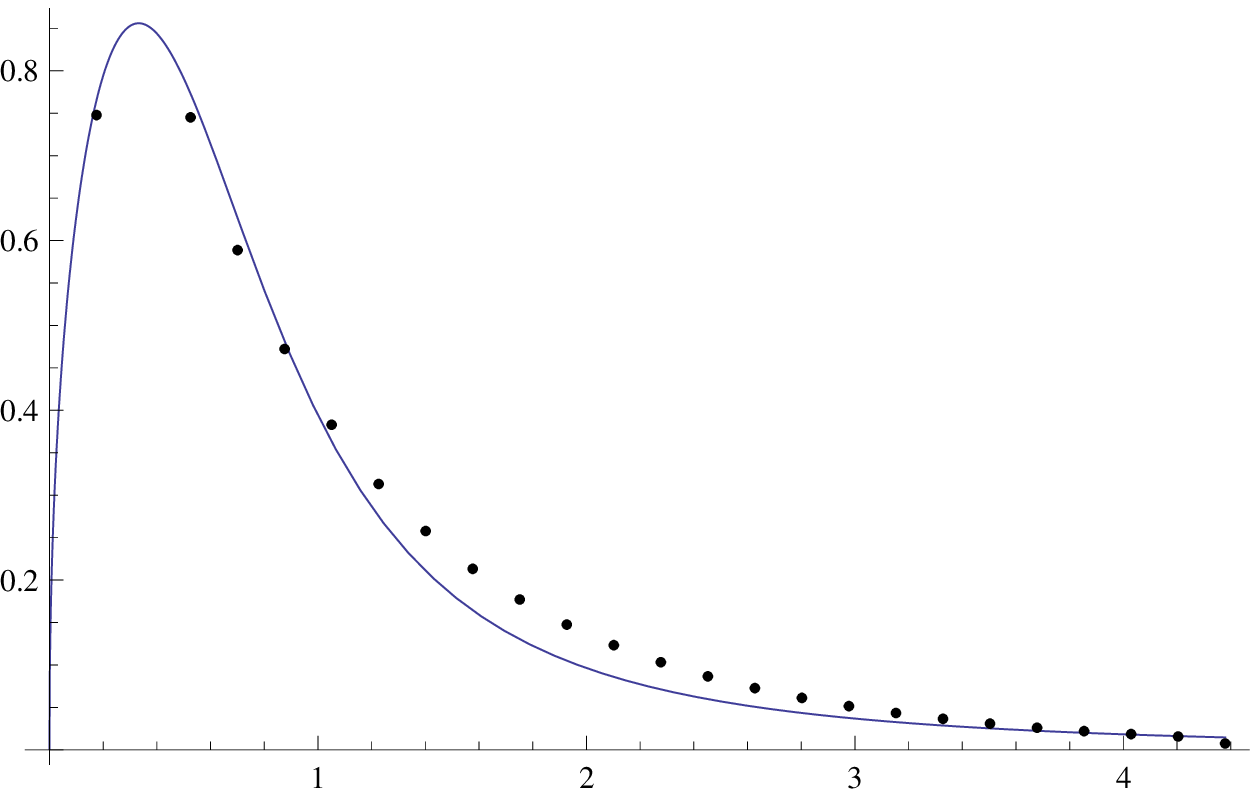}
 \end{center}
 \begin{center}
  \includegraphics[width=5.5cm,height=3cm]  {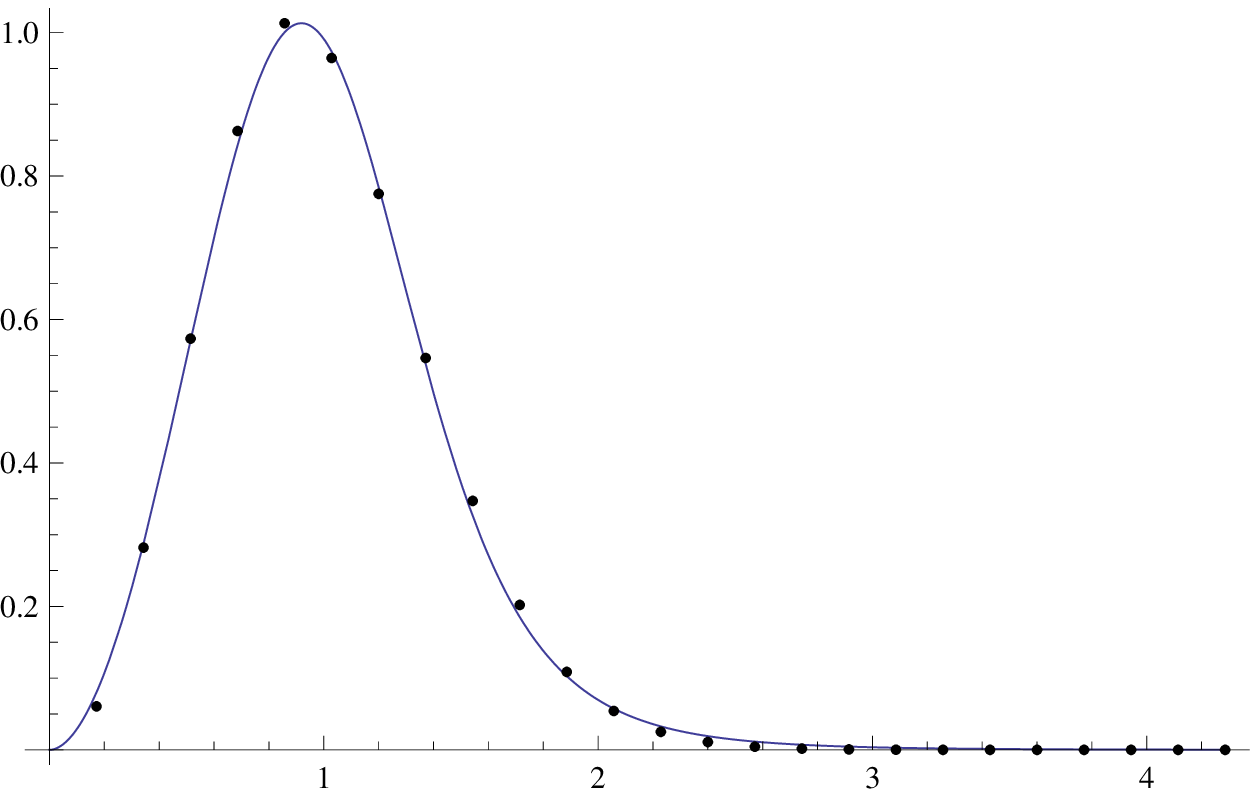}
  \hskip0.35cm
  \includegraphics[width=5.5cm,height=3cm]  {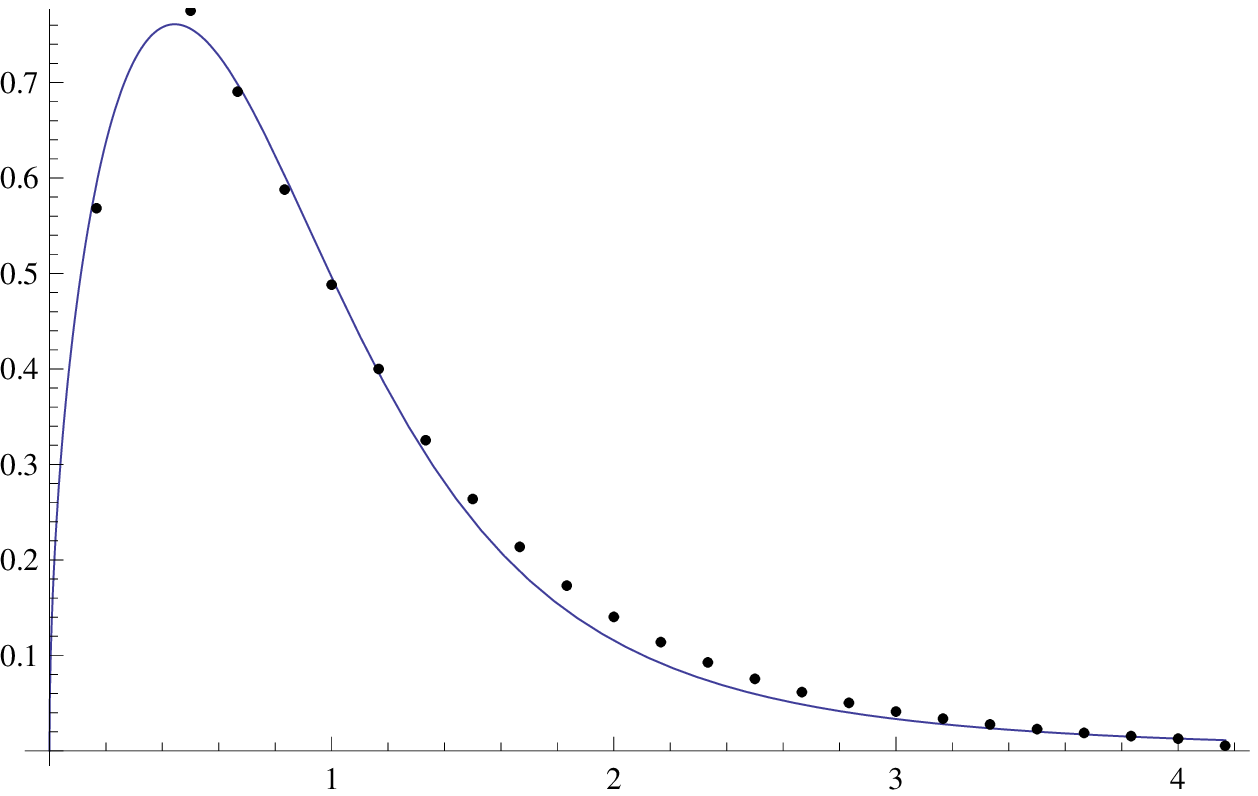}
   \hskip0.35cm
  \includegraphics[width=5.5cm,height=3cm]  {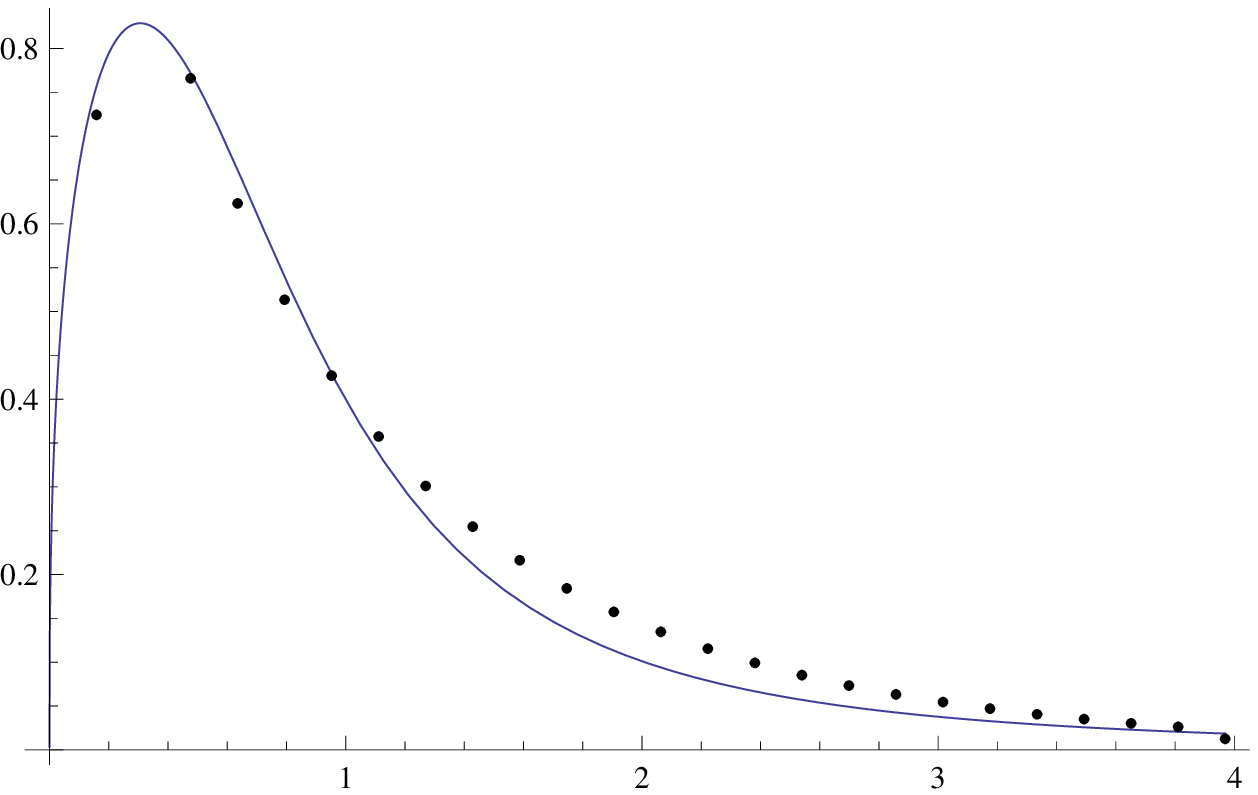} 
  \end{center}
\caption{In each of the two panel rows three fittings are illustrated
of three asymptotic stationary distributions relative to the three {\it{Cases}} 
{\it{FSP}}, {\it{TID}}, and {\it{FAE}} of Section \ref{diff cho} with the $\kappa$-generalized distribution.
The taxations rates $\tau_j$ are different in the simulations relative to the first and to the second row. 
The horizontal variable
is defined as 
$x = z/\langle z \rangle$, with $z$ the absolute personal income and $\langle z \rangle$ its mean value.
The vertical variable is rescaled so as to have a normalized distribution.
The parameters $\alpha, \beta$ e $\kappa$ of the $\kappa$-generalized distribution $(\ref{Kan distribution})$ are as follows.
In the first row:
{\it{FSP}}: $\alpha = 2.9459$, $\beta = 0.7754$, $\kappa = 0.4426$;
{\it{TID}}: $\alpha = 1.5113$, $\beta = 1.0574$, $\kappa = 0.6156$;
{\it{FAE}}: $\alpha = 1.4601$, $\beta = 1.3008$, $\kappa = 0.9952$;
in the second row:
{\it{FSP}}: $\alpha = 2.8868$, $\beta = 0.7776$, $\kappa = 0.4342$;
{\it{TID}}: $\alpha = 1.5118$, $\beta = 1.0560$, $\kappa = 0.6138$;
{\it{FAE}}: $\alpha = 1.3784$, $\beta = 1.2171$, $\kappa = 0.8664$.} 
\end{figure*}
%%%%%%%%%%%%%%%%%%%%%%%%%%%

The Amoroso distribution 
$$
f(x) = \frac{(\alpha-1)^\alpha}{\Gamma(\alpha)} \frac{e^{-\frac{\alpha-1}{x}}}{x^{1+\alpha}}
$$
contains one parameter and
is the exact limit of the model by Bouchaud and Mezard
\cite{BouMez}.
It is characterized by the presence of a maximum, by a null value
in zero (with exponential increase) and by a Pareto index $\alpha$ which
is independent from the average income. 

The Pareto-Gompertz
distribution contains three independent pa\-ram\-e\-ters, is mono\-ton\-ically
decreasing and has a Pareto index which depends on the average
income. Its analytical expression is more complex:
it is given by a combination of
a double exponential of the form
$f(x) = B e^{A-Bx} e^{e^{A-Bx}}$
followed by a power tail. 
It has been successfully employed
e.g. in \cite{ChaFigMouRib}
for the description of real statistical data
from Brazil.

Rational functions of the form
$f(x)=ax^\gamma/(1+bx)^{\gamma+\delta}$
are suggested by Scafetta, West and Picozzi in \cite{ScaWesPic} as
apparently well fittings
the computer-generated wealth distributions
of their trade-investment model.
 
\smallskip

The $\kappa$-generalized distribution, given by
\begin{equation}
f(x) = \frac{\alpha \beta x^{\alpha-1} exp_{\kappa}(- \beta x^{\alpha})}{\sqrt{1+\beta^2 \kappa^2 x^{2\alpha}}} \vb
\label{Kan distribution}
\end{equation}
with the generalized exponential function
$$
exp_{\kappa}(y) = (\sqrt{1 + \kappa^2 y^2} + \kappa y)^{1/\kappa}
$$
was introduced by Kaniadakis in \cite{Kan} in a general context.
In \cite{CleGalKan} and \cite{CleDiMGalKan} it has been considered in connection with 
its application to the economic problem of the representation of income distribution,
with the parameters
$\alpha, \beta > 0$ and $\kappa \in [0,1)$. 
In that context, the 
variable $x$ in $(\ref{Kan distribution})$
is defined as 
$x = z/\langle z \rangle$, with $z$ the absolute personal income and $\langle z \rangle$ its mean value.
One of the interesting properties that this probability density function enjoys is its behavior for large values of $x$;
indeed,
$$
f(x) \sim { \frac{\alpha}{\kappa} \, (2 \beta \, \kappa)^{-1/\kappa} } \, x^{- (\frac{\alpha}{\kappa} + 1)} \qquad \hbox{for} \ x \to + \infty \pb
$$
The parameters $\alpha$ and $\beta$ respectively determine
the curvature (shape) and the scale of the probability distribution.
The parameter $\kappa$ measures the fatness of the upper tail: the larger its magnitude is, the fatter the tail is.
The strength of the $\kappa$-generalized distribution stems from its
ability to reproduce in an excellent way empirical data of several countries 
(see \cite{CleGalKan} for data concerning Germany, Italy, United Kingdom
and
\cite{CleDiMGalKan} for
Australia, United States).

\smallskip

Based on
several attempts to fit the asymptotic stationary distributions 
resulting from
our simulations with one or the other of the mentioned analytical expressions,
we
are in the position of claiming that the unique among them
which seems to be suitable for the purpose
is the $\kappa$-generalized distribution. 
In fact,
it displays quite good performances.

The Figure $4$ illustrates in each of the two panel rows 
a sample of three fittings out of many,
relative to the three {\it{Cases}} {{\it{FSP}}, {\it{TID}} and {\it{FAE}}} introduced in Section \ref{diff cho}.
The
average incomes are $r_j= 10 j$.
The taxations rates $\tau_j$ are taken differently in the simulations relative to the first and to the second row. 
Specifically, 
in the simulations whose asymptotic trend is shown in the first row, three taxation rates are postulated, given by:
$$
\begin{array}{llll}
\tau_j = 5/100 \quad & \hbox{for} \ j = 1,...,5 \vb \nonumber \\
\tau_j = 25/100 \quad & \hbox{for} \ j = 6,...,20 \vb \nonumber \\
\tau_j = 45/100 \quad & \hbox{for} \ j = 21,...,25 \vb \nonumber 
\label{3Aliquote}
\end{array} 
$$
while in the simulations relative to the second row
the taxation rates are as in
$(\ref{progressivetaxrates})$.
In each of the plots the horizontal variable is given by
$x = z/\langle z \rangle$, with $z$ the absolute personal income and $\langle z \rangle$ its mean value
and
the vertical variable representing the fraction of individual with a certain income is rescaled so as 
to have a distribution.

%%%%%%%%% %%%%%%%%% %%%%%%%%% %%%%%%%%% %%%%%%%%%
\section{Some remarks on reversibility}
\label{remarks reversibility}
%%%%%%%%% %%%%%%%%% %%%%%%%%% %%%%%%%%% %%%%%%%%%

According to general results of kinetic theory \cite{Yak} any 
system with reversible microscopic interactions approaches,
at equilibrium, a Gibbs distribution function (a pure
exponential). The reversibility condition is satisfied when
the probability of each inverse elementary interaction process
in the system is equal to the probability of the corresponding
direct process.

%%%%%%%%%%%%%%%%%%%%%%%%%%%
\begin{figure}
  \begin{center}
   \resizebox{0.75\columnwidth}{!}{
   \includegraphics[width=8cm,height=8cm]  {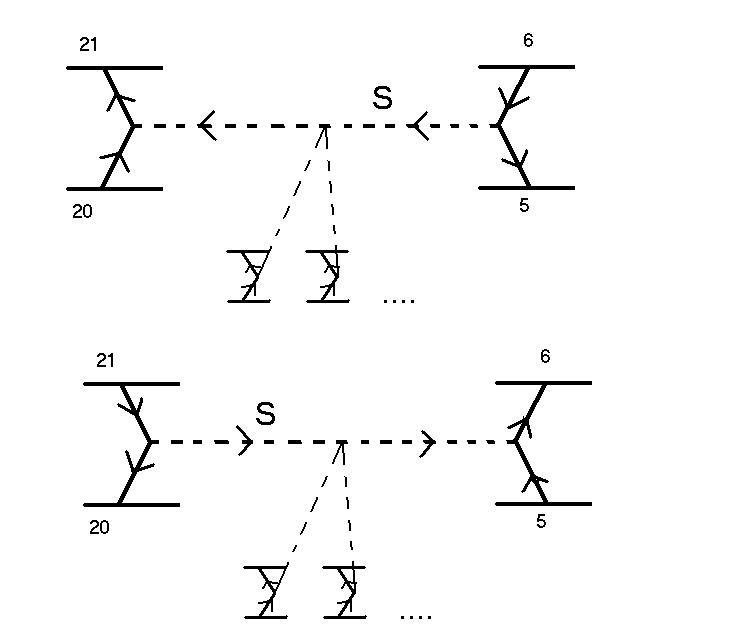}
   }
   \end{center}
\caption{Example of non-reversible process with taxes.
The first diagram  shows an example of an interaction between a 
\textquotedblleft rich\textquotedblright 
agent (belonging to the $20$-th income class) and a 
\textquotedblleft poor\textquotedblright 
agent ($6$-th class). In this case the poor agent pays an amount S and acquires a definite probability to pass to the lower class. 
The rich agent gains an amount $S$ minus the taxes and acquires a probability to pass to the higher class. 
The taxes are distributed among all the other agents, giving each of them a (small) probability of transition to their upper classes. 
The second diagram represents the inverse process, where the rich agent pays the poor agent and thus acquires a probability to pass to the lower class. 
Suppose that the two alternatives 
\textquotedblleft $6$ pays $20$\textquotedblright 
and 
\textquotedblleft $21$ pays $5$\textquotedblright 
have the same probability. If the rich and the poor are subjected to different taxation rates, 
then the two diagrams have different probabilities and the whole process is clearly non-reversible. 
But even if the taxation rate is the same, the whole process is non-reversible, 
because there exists no way for all the other agents to give the tax money back. In other words, the exchange of the amount $S$ 
can be reversed, but not the \textquotedblleft small re-distributions\textquotedblright  of the tax on S. 
}
\label{onecolumnreversibility}       
\end{figure}
%%%%%%%%%%%%%%%%%%%%%%%%%%%

For our model family, a typical example of elementary interaction
with money exchange and taxation is shown in Fig. 5 (see
detailed discussion below). This interaction is reversible
in only one case among those described in this paper,
and more precisely
only provided the following three conditions 
are satisfied (see Table $1$):

\begin{enumerate}
\item the parameters $b^k_{ij}$ are as in the {\it{Case FAE}} of the
model family, in which the exchanged amount is fixed and the saving
propensity is variable; 
\item the class income values $r_j$
are linear in the class index~$j$; 
\item the taxation rates
are set equal to zero. 
\end{enumerate}

Let us analyse these three conditions in
detail.

\smallskip

As for condition $1$, it is well known \cite{Yak} and easily realised that wealth
exchange processes with fixed saving propensity are non-reversible. Such 
processes are present in the {\it{Case FSP}} of
our model, where $p_{hk}=(1/4) \, r_h/r_n$. In {\it{Case TID}},
where $p_{hk}=(1/4) \, \min\{r_h/r_k\}/r_n$, the interaction is also
non-reversible, because the quantity $\min\{r_h/r_k\}$ is in
general different in the direct process and in the inverse
process. 

In connection with condition $2$, we notice that the income differences $|r_i-r_{i-1}|$ in the denominators
of the parameters $b_{hk}^i$ in $(\ref{bhk})$
are independent from the index $i$
only if $r_i$ is linear in $i$. Otherwise, the income differences
depend on income, being typically larger for the rich
classes. It follows that also in that case the probabilities
of the direct and inverse wealth exchange processes are
not the same.

Finally, about condition $3$, even assuming that the conditions conditions $1$ and $2$ hold true, we 
have irreversibility due to the taxation process.
This can be grasped by looking at 
the Figure $5$, which is supposed to refer 
to {\it{Case FAE}} of our model family, with linear class incomes.
The first diagram shows an example of an interaction 
between a 
\textquotedblleft rich\textquotedblright 
agent (belonging to the $20$-th income class) and a 
\textquotedblleft poor\textquotedblright 
agent ($6$-th class). In this case the poor agent pays an amount $S$ 
and acquires a definite probability to pass to the lower class. 
The rich agent gains an amount $S$ minus the taxes and acquires a 
probability to pass to the higher class. 
The taxes are distributed among all the other agents, giving each 
of them a (small) probability of transition to their upper classes. 
The second diagram represents the inverse process, where the rich 
agent pays the poor agent and thus acquires a probability to pass 
to the lower class. 
Plainly, if the rich and the poor 
are subjected to different taxation rates, then the two diagrams have 
different probabilities and the whole process is non-reversible. 
But even if the taxation rate is the same, the whole process is non-reversible, 
because there exists no way for all the other agents to give the tax money back. 
In other words, the exchange of the amount $S$ can be reversed, but not 
the \textquotedblleft small redistributions\textquotedblright \ of the tax on~$S$. 

\smallskip

Hence, 
calculating the distribution function of
our mo\-del in {\it{Case FAE}}, with linear income classes and
no taxes, provides an important cross-check of the mo\-del and of
the numerical solutions: we expect in this case to obtain
the pure Gibbs distribution predicted by kinetic theory
for reversible processes. And this is, actually, what is
found: one obtains an exponential distribution, except for
the first and last classes, whose wealth exchanges are non-reversible due to boundary effects.

In this connection we also recall that the initial conditions
play in general a role in the equilibrium approach to a
Gibbs distribution: such an equilibrium distribution can be
reached in a conservative system only if the initial conditions do not display
a population inversion corresponding to negative temperature.
In our case this is guaranteed by the choice of an initial
population concentrated in the classes with low income 
(compare Section \ref{diff cho}).

The emergence, in the reversible case without ta\-xes, of
a purely exponential distribution is at the same time natural
and puzzling. One can naturally expect, in fact, that the removal
of taxes causes an increase in the population of the poorest
at the detriment of the middle classes, 
as it happens in the exponential distribution.
On the other hand, the exponential tail is thinner than the
power law tail of the (non-reversible) distribution with
taxes, and the reason for this is less obvious. Things go
as if the complete reversibility of the wealth exchange
would make it harder for the super-rich to preserve their
wealth.

We close this section recalling that we pointed out in Section \ref{diff cho}
a similarity between the {\it{Case FAE}} 
(with a fixed exchanged amount of money)
of our model family and some models without taxes and
with variable saving propensity (\cite{ChattChakrManna}, 
see also
\cite{PatHeinCha} and
\cite{MatTos}). We
observed that our fixed exchanged amount of money mimics 
a saving propensity which varies among agents belonging
to different classes
and that in both cases a Pareto tail is
found. Since, however, the removal of the taxes from our
model causes the Pareto tail to disappear, one should
conclude that the dynamical mechanism behind the formation
of the tail is different in the two cases.

%%%%%%%%%%%%%%%%%%%%%%%%%%%
\begin{table}
\caption{Summary of results. The three basic model cases considered in this paper (first column) 
and their further subdivision/specialization according to taxation and 
linear/nonlinear 
class incomes. The last two columns record the appearence of reversibility and power law tails. 
Notice that linear class incomes in general do not allow to distinguish clearly between a power law tail and an exponential tail.}
\label{tab:1}       % Give a unique label
%For LaTeX tables use
%\begin{tabular}{lll}
%\hline\noalign{\smallskip}
\smallskip
\centering % centering table
\begin{tabular}{l c c c} % creating 4 columns
\hline \hline % inserting double-line
Model & Income & Reversibility &Fat tail \\ [0.5ex]
\hline \hline % inserts double-line
% Entering 1st row
&linear &no & ?\\ [-1ex]
\raisebox{1.5ex}{{\it{FSP}} with taxes} &non-lin.
& no & no \\[1ex]
\hline % inserts single-line
% Entering 2nd row
&linear &no & ?\\ [-1ex]
\raisebox{1.5ex}{{\it{FSP}}, no taxes} &non-lin.
& no & no \\[1ex]
\hline % inserts single-line
% Entering 3rd row
&linear &no & ?\\ [-1ex]
\raisebox{1.5ex}{{\it{TID}} with taxes} &non-lin.
& no & yes \\ [1ex]
\hline % inserts single-line
% Entering 4th row
&linear &no & ?\\ [-1ex]
\raisebox{1.5ex}{{\it{TID}}, no taxes} &non-lin.
& no & yes \\[ 1ex]
\hline % inserts single-line
% Entering 5th row
&linear &no & ?\\[ -1ex]
\raisebox{1.5ex}{{\it{FAE}} with taxes} &non-lin.
& no & yes \\[1ex]
\hline % inserts single-line
% Entering 6th row
&linear &yes & no\\[ -1ex]
\raisebox{1.5ex}{{\it{FAE}}, no taxes} &non-lin.
& no & yes \\[1ex]
\hline % inserts single-line
%\label{tab:PPer}
\end{tabular}
% Or use
%\vspace*{5cm}  % with the correct table height
\end{table}
%%%%%%%%%%%%%%%%%%%%%%%%%%%

%%%%%%%%% %%%%%%%%% %%%%%%%%% %%%%%%%%% %%%%%%%%%
\section{Concluding remarks}
\label{conclusions}
%%%%%%%%% %%%%%%%%% %%%%%%%%% %%%%%%%%% %%%%%%%%%

The aim of this paper is to provide
further insight into the
formation process of stationary income profiles. 
A family of models introduced in \cite{Ber} and \cite{BerMod}
has been investigated.
In particular, the high flexibility of these models (related to the wide range of possible parameter choices) has been exploited
for the construction of various examples.
A comparison between different long time behaviors
exhibited by those examples
contributes to further understanding of the mechanisms responsible for 
the emergence of
one or the other output.
Attention has also been given to placing our results 
within a frame of related results obtained through 
different methods.

Our approach for the description of the process of wealth exchange in the presence of taxation 
comprises
a subdivision of the population into income classes, and
a corresponding discretization of the distribution function.
This approach has proven to be powerful in its
application. It translates well-tested physical methods
of kinetic theory into a manageable mathematical framework
(a system of ordinary differential equations) which can be
quickly solved numerically with standard software.
The microscopic interaction parameters 
as well as the taxation rates
have a straightforward
interpretation and can be easily varied. This allows to
reproduce the qualitative results of several different models. 

In particular, in correspondence to some parameter choices,
stationary distributions have been found, whose 
higher income sections exhibit
a power law decreasing behavior.
It could be argued in this connection
that properly talking of power law tails requires treating the income or wealth variable 
as a continuous one taking values in $[0,+\infty)$.
This is for example what is done in papers like \cite{DurMatTos} and
\cite{MatTos}, where a rigorous analytical approach is developed, which
includes the study of the evolution of the moments of the solution.
Certainly, admitting a finite number of income classes entails 
some unnatural constraint and causes some boundary effect as well.
On the other hand, we think that
for the application-oriented problem at issue this is not so out of place:
typically, in reports and studies concerning real world populations,
income distributions are never infinitely extended,
a finite number of income bins are taken into consideration,
and of course, 
statements concerning the power law decreasing behavior 
are not to be intended in a strictly analytical way.
We rather observe that for a more credible depiction
of modern market economies,
more relevant corrections should be incorporated in the model,
the first one being probably the introduction of a term accounting for a possible generation
of wealth due to production efficiency increase.
In fact, the conservation of the total wealth is not a realistic assumption.

We are very far from thinking and claiming
that the models investigated here 
can be employed to explain in a satisfactory way real world situations.
However, they may provide a starting point for further progress in this direction.
The reason for their interest, we think, is given by
their flexibility and ability of reproducing the emergence of qualitative global features
based only on the description of the micro-level interactions.
And this is one of the main challenges in a complex systems approach.

\end{document}